\documentclass[useAMS,usenatbib]{mn2e}
\usepackage{graphicx}
\usepackage{txfonts}
\usepackage{natbib}

\usepackage[english]{babel}
\usepackage{color}
\DeclareGraphicsExtensions{.pdf,.ps,.jpg}
\title[The space density of z$>$4 blazars]{The space density of z$>$4 blazars}
\author[Caccianiga et al.]{A. Caccianiga$^1$, A. Moretti$^1$, S. Belladitta$^{1,2}$, R. Della~Ceca$^1$, S. Ant\'on$^{3}$, L. Ballo$^4$, 
\newauthor C. Cicone$^1$, D. Dallacasa$^{5,6}$, A. Gargiulo$^{7}$, 
L. Ighina,$^{1,8}$, M.J. March\~a$^9$, P. Severgnini$^1$ 
\vspace{0.2cm}
\\
$^1$INAF - Osservatorio Astronomico di Brera, via Brera 28,  I-20121 Milan, Italy\\
$^2$DiSAT, Universit\`a degli Studi dell'Insubria, Via Valleggio 11, 22100, Como, Italy\\
$^3$CIDMA, Department of Physics, University of Aveiro, 3810-193, Aveiro, Portugal\\
$^4$XMM-Newton Science Operations Centre, ESAC/ESA, PO Box 78, E-28691 Villanueva de la Ca\~ada, 
Madrid, Spain 0000-0002-5036-3497\\
$^5$Dipartimento di Astronomia, Universit\`a di Bologna, via Ranzani 1, 40127, Bologna, Italy\\
$^6$INAF - Istituto di Radioastronomia, Via Gobetti 101, I-40129 Bologna, Italy\\
$^7$INAF - Istituto di Astrofisica Spaziale e Fisica Cosmica (IASF), Via E. Bassini 15, I-20133 Milano, Italy\\
$^8$Dipartimento di Fisica G. Occhialini, Universit\`a di Milano-Bicocca, Piazza della Scienza 3, I-20126 Milano, Italy\\
$^9$Department of Physics \& Astronomy, University College London, Gower Street, London, WC1E 6BT, UK
 }

   \date{}

\begin{document}

\label{firstpage}

\maketitle

\begin{abstract}
High redshift blazars are an important class of Active Galactic Nuclei (AGN) that can provide an independent estimate of the supermassive
black-hole mass function in high redshift radio-loud AGN without the bias due to absorption along the line-of-sight.
Using the Cosmic Lens All Sky Survey (CLASS) we built a complete radio flux-limited sample of high redshift (z$>$4) blazars  suitable for statistical studies. By combining dedicated optical  observations and the SDSS spectroscopic database, we obtained a sample of 26 blazar candidates with a spectroscopic redshift above 4. On the basis of their radio spectrum we distinguish between blazars and QSO with a Gigahertz Peaked Spectrum (GPS) like spectrum. Out of the 18 confirmed  blazars  14 constitute a completely identified, flux-limited 
sample down to a magnitude of 21 (AB).
Using this complete sample we derive a space 
density of blazars with 4$<$z$<$5.5 of $\rho$=0.13$^{+0.05}_{-0.03}$  Gpc$^{-3}$.  This is the first actual estimate of the
blazar space density in this range of redshift. 
This value is in good agreement with the extrapolation of the luminosity function and cosmological
evolution based on a sample of flat-spectrum radio quasars selected at lower redshifts and
it is consistent with a cosmological evolution peaking at z$\sim$2 similar to radio-quiet QSO. 
We do not confirm, instead, the presence of a peak at z$\sim$4 in the space density evolution, recently suggested using an X-ray selected sample of blazars. 
It is possible that this extreme peak of the evolution is present only among the most luminous blazars.

\end{abstract}


   \maketitle

%__________________________________________________________________

\section{Introduction}
Supermassive black holes (SMBH) are the result of the evolution of the mass inflow
towards the nuclear region of galaxies across cosmic time (\citealt{Merloni2015} 
for a recent review). 
A fundamental question is how the observed local population of SMBH can be traced
back to the original seeds through different phases of growth, identified by the
intense nuclear accretion activity (AGN). High redshift AGN surveys represent the
best tool to provide observational constraints to current theoretical models of
SMBH formation (e.g. \citealt{Volonteri2010}; \citealt{Valiante2017}).  
New surveys planned for the next decades will play a 
fundamental role in this field by selecting 
thousands of high-z AGN.  However, obscuration from circumnuclear matter 
is expected to introduce selection biases that are difficult to
quantify (see e.g. \citealt{McGreer2006}; \citealt{Zeimann2011}; \citealt{Vito2018}). 
Radio selections, like those that will be carried out
with Square Kilometre Array (SKA) or with its precursors (e.g. see \citealt{Norris2011a}), 
can provide a less biased census of high-z 
AGN (at least of the radio-loud fraction) since radio wavelengths are 
unaffected by absorption. This is true, however, only if the selection of the
high-z radio sources is not based on the colours of the optical counterparts otherwise
the usual biases against absorbed AGN will be present also in these samples.

A complementary approach  has been recently proposed (\citealt{Volonteri2011}; \citealt{Ghisellini2014}; \citealt{Sbarrato2014a}) based on the class of 
blazars, i.e. radio-loud
AGN whose relativistic jet points towards the observer  (\citealt{Urry1995}).
The particular orientation of these sources makes obscuration less important, since the jet is
expected to clear out the path along the line-of-sight to the active nucleus and the dusty torus should not intercept
the radiation.
Another advantage of studying blazars is that their compact radio emission is less affected by the 
attenuation due to the interaction of radio photons with the Cosmic Microwave Background (CMB)
if compared to the mis-aligned, lobe-dominated radio-loud AGN (e.g. see \citealt{Ghisellini2015d},
\citealt{Wu2017}). 
The space density of the entire population of radio-loud AGN
can then be inferred indirectly from the observed number of  
blazars since we expect: N$_{total}$=N$_{blazars}\times 2 \Gamma^2$, where 
$\Gamma$ is the Lorentz factor of the bulk velocity in the jet (typically, 
$\Gamma\sim$10-15, e.g. \citealt{Ghisellini2014}). 

The use of blazars to study the high redshift Universe requires the
selection of  well-defined and sizable samples suitable for reliable statistical analyses. 
\citet{Sbarrato2013a} used a systematic approach
to select a well-defined sample of high-z blazars starting
from the z$>$4 QSO present in the SDSS (DR7) and
considering only the most radio-loud ones that are
expected to be mostly blazars. The
recognition of the actual blazars in the sample is
in progress. With this approach, \citet{Sbarrato2013a} are building up an {\it optically
selected} sample of blazars whose completeness is related to the actual completeness of the starting
data set (the SDSS DR7 spectroscopic database). 

Here we present an alternative (and complementary) approach 
aimed at building a {\it radio-flux limited} 
sample of high-z (z$>$4) blazars to be directly
comparable with radio samples selected at lower redshifts.
We do not limit our search to the SDSS sky area and to the
already available spectroscopic identifications.
The starting point is the CLASS (Cosmic Lens All Sky Survey, {\citealt{Myers2003}; 
\citealt{Browne2003}), a flux density limited survey at 5~GHz
(S$_{5GHz}>$30 mJy) covering almost the entire northern hemisphere ($\sim$16300 deg$^2$) 
and that contains $\sim$11,000 flat spectrum radio sources. 
The  goal is to create the largest radio-selected sample of {\it bona-fide} high-z blazars
to be used for statistically-sound studies.
Since blazars have flat radio spectra, this survey
represents the ideal starting point for an efficient selection. CLASS is also deep 
enough to allow the detection of powerful ($\sim$10$^{27}$ W Hz$^{-1}$) blazars at z=5-6. 
Using existing spectra (mostly from SDSS) {\it plus} dedicated observing runs at 4-10 meter class
telescopes we have
now completed (at 95\% level) the identification of the sample down to a well defined
optical magnitude (21) in a filter corresponding to a rest-frame $\sim$1200-1400\AA\ wavelength range (i.e. either {\it r}, {\it i} or {\it z} filter depending on the  redshift of the source.)

In this paper we present this sample together with a first
estimate of the space density of blazars in the 4-5.5 redshift bin.
In Section~2 we discuss how the sample has been selected while in Section~3 we present the
spectroscopic follow-up that we have recently carried out to complete the
identification of the high-z candidates.
The completeness of the sample is studied in Section~4. We then use (Section~5) the radio spectral
indices over a wide range
of frequencies, from 150~MHz up to 8.4~GHz, to distinguish between genuine
flat-spectrum radio sources 
from steep spectrum objects or sources with a peaked spectrum. We finally derive the
volume density of blazars versus redshift (between 4 and 5.5, Section~6) and compare
it with the predictions based on samples selected at lower redshifts (Section~7). 
Conclusions are reported in Section~8.

Throughout the paper we assume a flat $\Lambda$CDM cosmology with H$_0$=70 km 
s$^{-1}$ Mpc$^{-1}$, $\Omega_{\Lambda}$=0.7 and $\Omega_{M}$=0.3.  Spectral 
indices are given assuming S$_{\nu}\propto\nu^{-\alpha}$.

%______________________________________________ Gamma_1 (lg rho, lg e)
   \begin{figure*}
   \centering
    \includegraphics[width=15cm, angle=0]{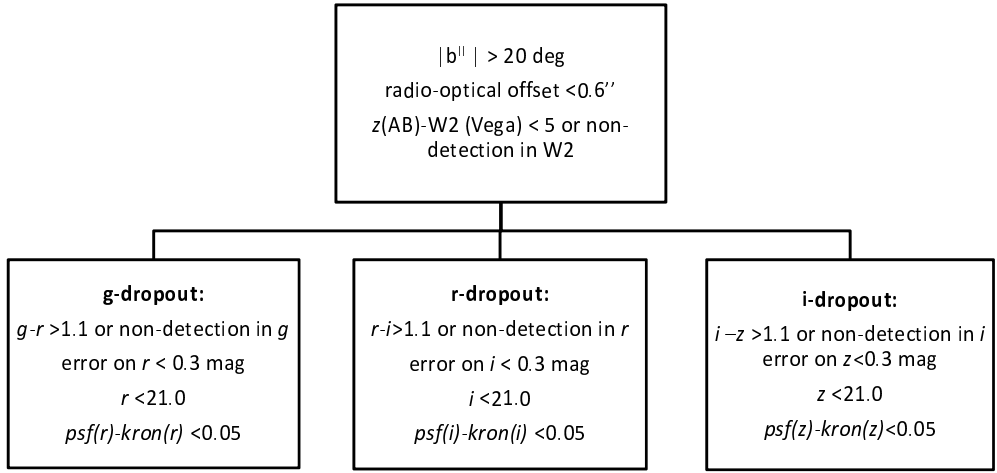}
   \caption{Summary of the criteria adopted to select high-z candidates from the
CLASS survey (see text for details)}
              \label{selection_flow}
    \end{figure*}
%-------------------------------------------------

\section{The CLASS sample of high-z AGN candidates}
CLASS is a flux limit survey at 5~GHz of 
flat spectrum radio sources, covering most of the northern sky (16300 deg$^2$). It was built by combining
the NRAO VLA Sky Survey (NVSS), at 1.4~GHz (\citealt{Condon1998}), with the Green-Bank Survey, GB6, at 5~GHz
(\citealt{Gregory1996}) and by selecting only the objects with a flat spectrum between 1.4 and 5 GHz 
($\alpha<$0.5,
with f$_\nu\propto\nu^{-\alpha}$). All the selected sources have been then followed-up at 8.4~GHz using the Very Large Array (VLA) in its most extended configuration (A-array) providing a
resolution of $\sim$0.2$\arcsec$. The VLA follow-up provided not only a flux density at 
relatively high frequency but also a very accurate (sub-arcsec) radio position. 
CLASS represents the ideal starting point for an efficient selection since blazars have flat radio spectra.  
In addition, the availability of radio positions with 
uncertainties below a fraction of arcsecond 
guarantees the detection of the correct counterpart, even at faint optical magnitudes, 
without significant spurious contamination. In particular, we use a search radius
of 0.6$\arcsec$ to find the optical counterpart of the CLASS sources. 

%______________________________________________ Gamma_1 (lg rho, lg e)
   \begin{figure*}
   \centering
\includegraphics[width=7cm,angle=0]{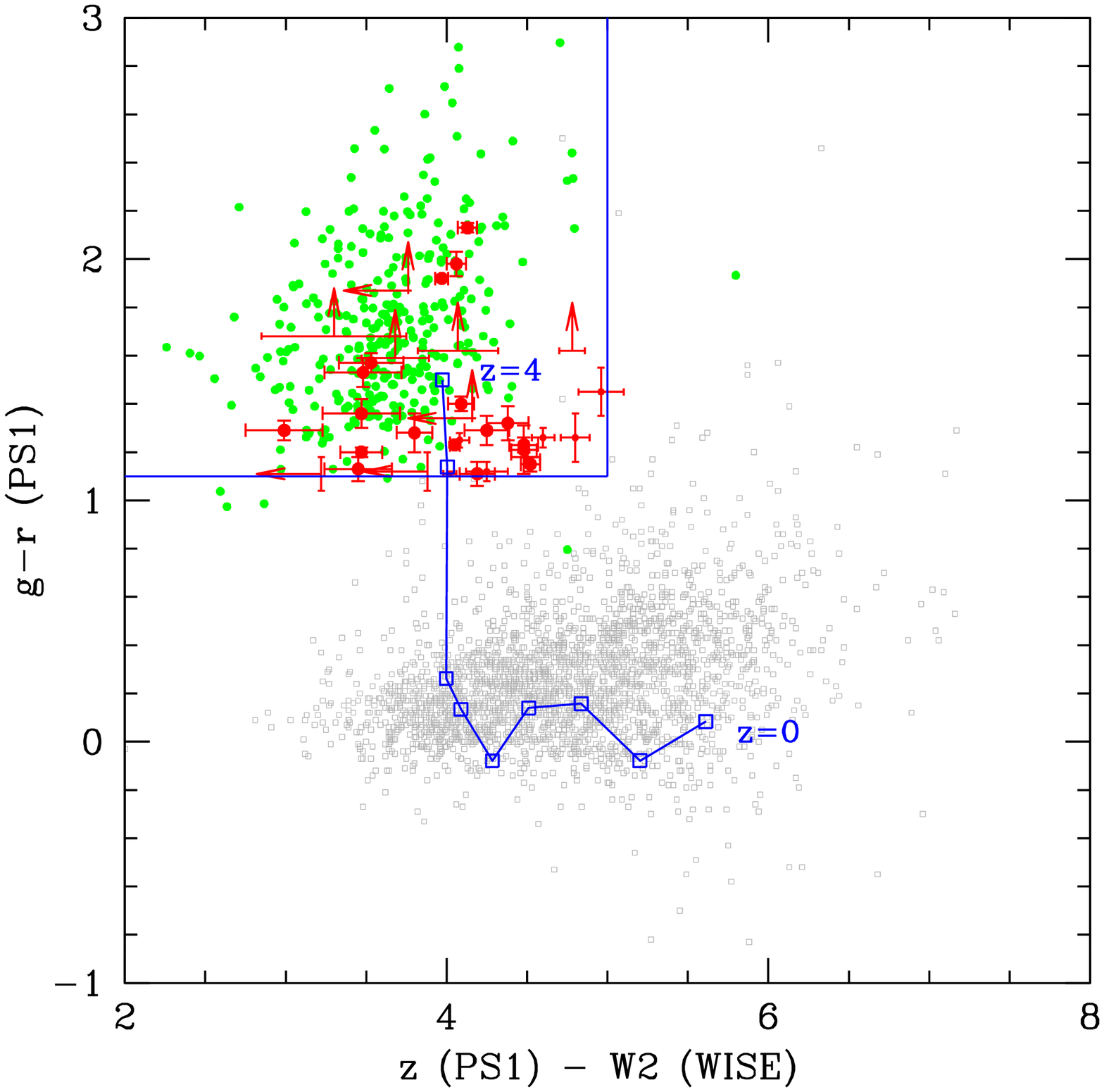}
\includegraphics[width=7cm,angle=0]{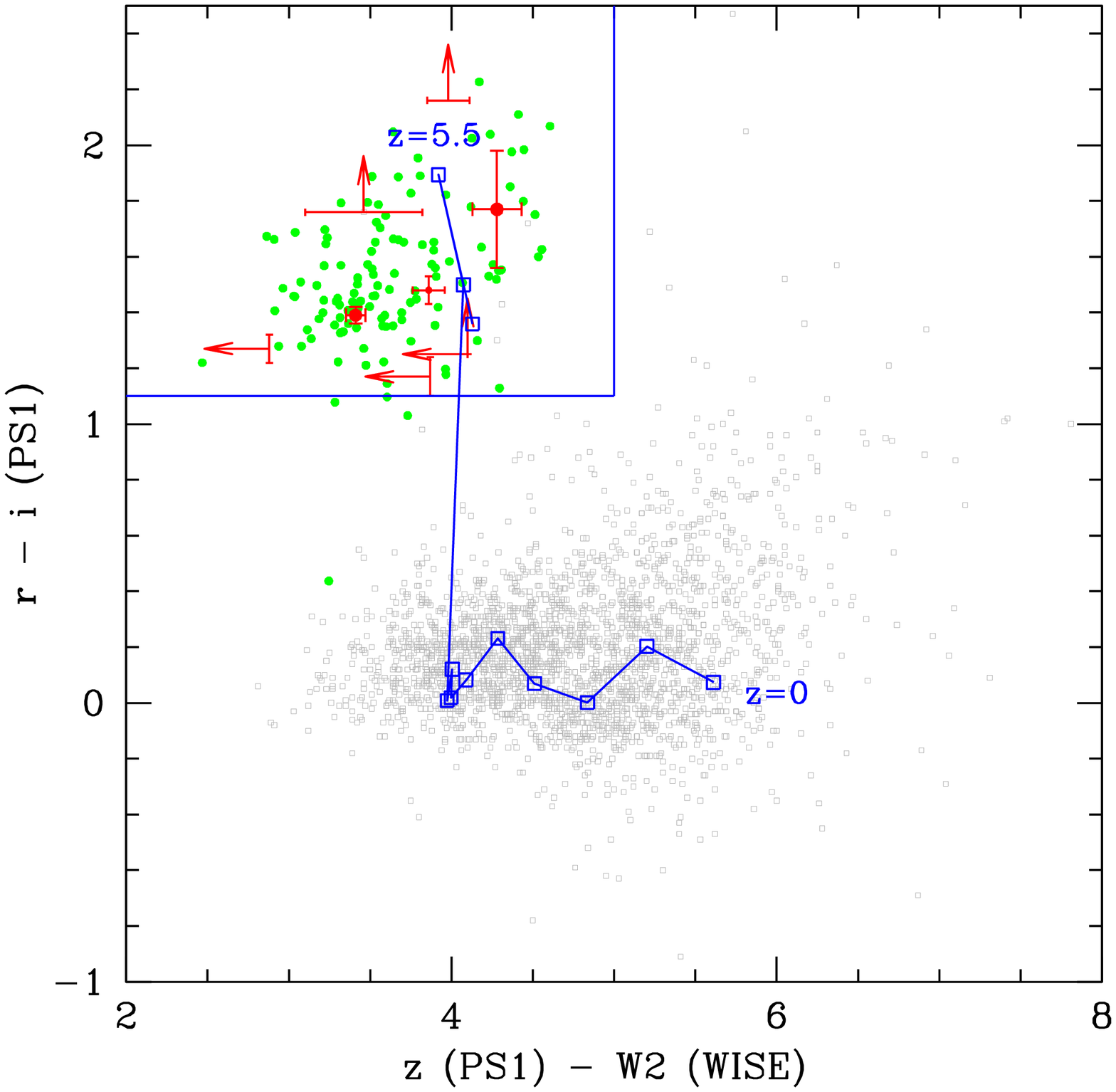}
\includegraphics[width=7cm,angle=0]{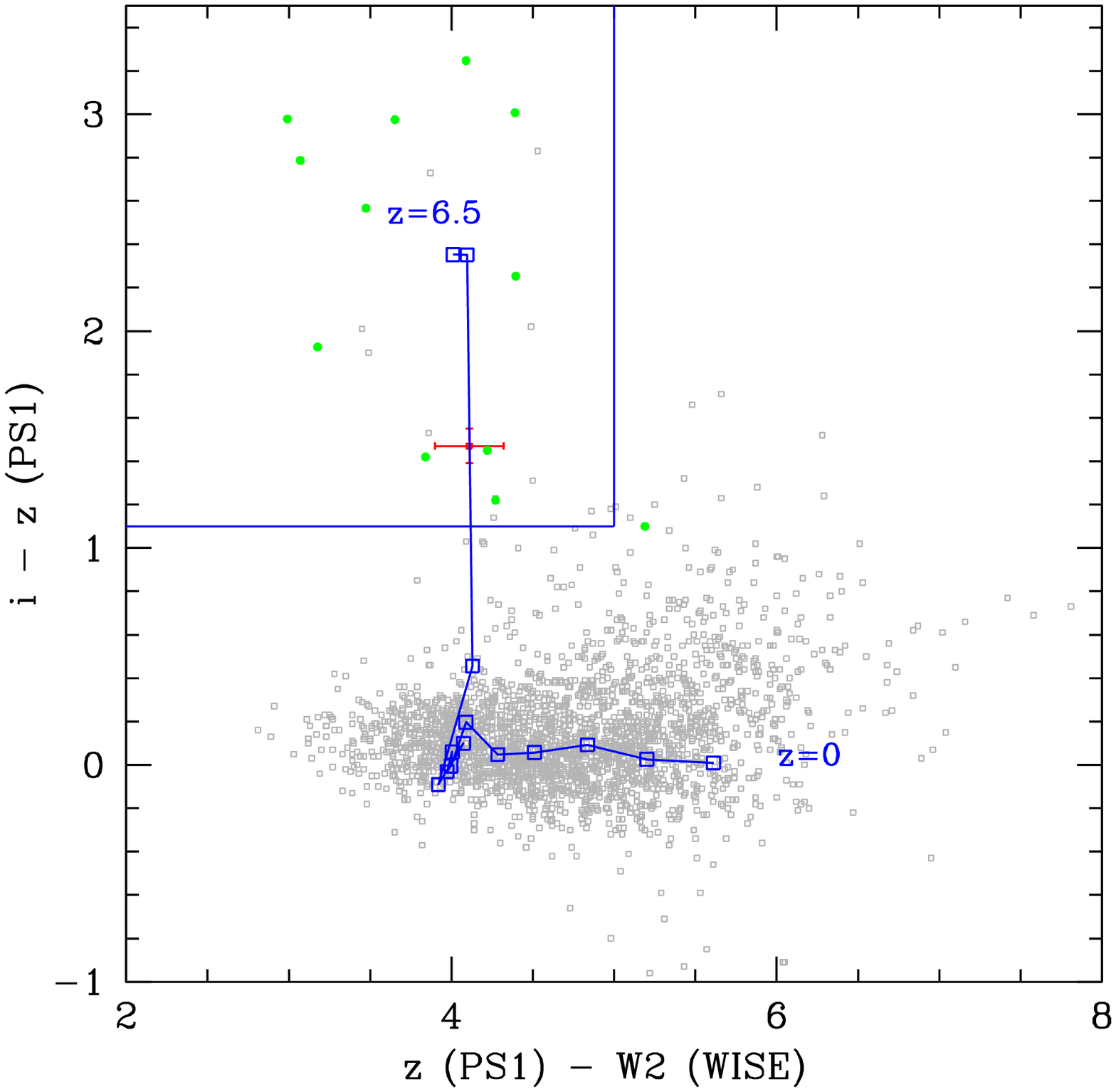}
\caption{Diagnostic diagrams used to select z$>$4 candidates from CLASS. We plot three different PS1 colours 
({\it g-r}, {\it r-i} and {\it i-z}) versus mid-IR colours 
({\it z}-W2), computed between PS1 {\it z}-filter and W2 filter of WISE at 4.6 micron (note that
magnitudes from PS1 are in AB systems while W2 magnitudes are in Vega system).
Red points are the CLASS objects that fulfill the selection criteria. 
Grey points show all the CLASS sources with a star-like PS1 counterpart and detected 
in WISE. Green-points are the high-z quasars from the literature
(including radio-quiet ones).
In particular, they are quasars at 4$<z<$4.5, at 4.6$<z<$5.5 and at 5.6$<z<$6.5 respectively in the three  panels. For clarity, error-bars are plotted only for the selected high-z candidates. For reference, we also plot the paths (blue lines) of the QSO template from \citet{Polletta2007} for redshifts ranging from 0 up to the maximum values indicated in each
panel by steps of 0.5. The template has been absorbed for the neutral hydrogen scattering in the inter-galactic medium along the line-of-sight at z=4, 4.5 and 5, respectively in the 3 panels, using the transmission curve from \citet{Songaila2004}.
Finally, the blue boxes indicate the regions used to
select the candidates. 
} 
              \label{diagnostic_plot}
    \end{figure*}
%-------------------------------------------------

First searches for high-z AGN from the CLASS survey started more than 15 years ago 
using the relatively shallow optical data available at that time (e.g. \citealt{Snellen2001}). 
Thanks to the recent release of the Panoramic Survey Telescope and Rapid Response System 
(Pan-STARRS1, PS1) survey (\citealt{Chambers2016}), we have now an almost complete
coverage of relatively deep photometric data in five bands ({\it g, r, i, z} and \it Y}).
Besides PS1 data we have also a complete coverage at mid-Infrared wavelengths from the Wide-field Infrared Survey Explorer (WISE, \citealt{Wright2010}).
We use all these photometric data to select high-z (z$>$4) AGN candidates in CLASS through the 
detection of the drop in the continuum flux when the Ly$\alpha$ forest and the Lyman limit systems 
enter a photometric filter ({\it dropout} method). 
This effect produces a very red colour between two filters, typically {\it g$-$r}, {\it r$-$i} or {\it i$-$z} (for the range of
redshift between 4 and $\sim$6, considered here). In order to obtain the highest level
of completeness we require that the
{\it g$-$r}, {\it r$-$i} or {\it i$-$z} are larger than 1.1. This threshold is suggested by the expected colours of a typical QSO template (see  Fig.~\ref{diagnostic_plot}) and it has been set so as to include  the large majority ($>$99\%) of the z$>$4 QSO discovered to date, as shown in Fig.~\ref{diagnostic_plot}.

A common contaminant for this type of selection are the dwarf stars. 
Usually, to minimize the inclusion of these contaminants it is often imposed an additional 
constraint on the colours of the source. For instance, for a {\it g}-dropout object
it is usually required a blue color between r and i-filters, while for an {\it r}-dropout
source it is required a blue {\it i-z} (e.g. see \citealt{Fan2000}, 
\citealt{Fan2003}).
However, this can introduce some 
incompleteness in the selection.
Therefore, we 
decided not to apply such a constraint also supported by the fact that a radio
selection excludes, by definition, the contamination from stars. 
The most important type of contamination in the case of a radio selected
sample are low redshift (z$\sim$1-2) ``red'' (i.e. moderately absorbed) AGN. 
This contamination can be kept at
minimum using mid-IR data from WISE since  
low-redshift red AGN typically have stronger $\sim$3 micron (observer's frame) 
emission (W2 magnitude) than high redshift objects with the same magnitude in the {\it z}-filter.
In particular, we impose that
the object is either undetected in WISE (W2) or, if detected, we require that
{\it z}$-$W2$<$5\footnote{we use the PSF magnitudes provided by PS1, that are in AB system, and
the magnitudes from allwise catalogue, that are in Vega system}. This threshold, which is very similar to the one
adopted by \citet{Carnall2015} to
search for high redshift QSO ({\it z}$-$W2$<$4.5), includes virtually all the known high-z QSO (see Fig.~\ref{diagnostic_plot})

Finally, we require that the source is star-like 
(psf magnitude$-$kron magnitude$<$0.05, see Chambers et al. 2016) and that the magnitude
in the reddest filter of each selection (i.e. {\it r} magnitude for {\it g$-$r} selection,
the {\it i} magnitude for the {\it r$-$i} selection and the {\it z} magnitude for the {\it i$-$z} selection) 
is brighter than 21. This last constraint guarantees that the dropout is significant
when the object is not detected in the bluest filter. The fact that the magnitude limit
(AB system) is the same in the three selections simplifies the statistical analysis of the sample since
it corresponds to a uniform threshold in the rest-frame wavelength interval of $\sim$1200-1400\AA\ for all the selected objects.

To reduce the number of possible spurious optical counterparts we further restricted the
search area to the high Galactic latitudes ($\mid b^{II}\mid \geq$20$\deg$).
The adopted selection criteria are summarized in Fig.~\ref{selection_flow}.

After this selection we inspected the high-z candidates and removed all the 
problematic cases (i.e. sources close to bright stars and/or with photometric 
problems). 
The complete list of high-z candidates selected in CLASS is presented in Tab.~\ref{dropout_sample}.
This list includes 37 objects out of which 25 have either a 
spectroscopic redshift from the literature or an available optical spectrum in 
the Sloan Digital Sky Survey (SDSS) catalogue. 
Nearly all the objects with an identification or an optical spectrum from SDSS are
high-z QSO with z above $\sim$3 except for 1 BL Lac object (GB6J154929+170853) that
is probably at z$\sim$1.2, based on a tentative detection of MgII$\lambda$2798\AA\  (z$\sim$1.25) and CIII]$\lambda$1909\AA\ (z$\sim$1.16){\footnote{SDSS reports z=0.62, in DR13 and DR14, and z=3.39 in DR12. None of these tentative 
redshifts have convincing  correspondences with the possible features observed in the spectrum and, therefore, we consider them unreliable.}.
In any case, the  optical spectrum of GB6J154929+170853  is
clearly ``blue'' with no obvious signs of dropout. In the SDSS (DR13) catalogue
the {\it g$-$r} is much lower ($\sim$0.4 mag) than what is reported in the PS1 
catalogue ($>$4). This is likely a case of problematic photometry in the PS1.

The remaining 12 selected candidates needed a spectroscopic follow-up.
To this end,  we carried out dedicated observations at {\it Large Binocular Telescope (LBT)} and at the {\it Telescopio Nazionale Galileo (TNG)} as detailed in the following section.

%%---------------------------------------------------------------------------------------------
\begin{table*}
\caption{The list of dropout candidates selected from CLASS}
\label{dropout_sample}
\begin{tabular}{c c c c c c r r r}
\hline\hline
 name            & CLASS &redshift & coordinates       & mag   & dropout & drop value & z$-$W2 & Ref \\ 
                 &       &    &   (J2000.0)       & (mag) &         & (mag)      &  (mag) &         \\
\hline

GB6J001115+144608 & QSO & 4.96 & 00 11 15.24 +14 46 01.8 & 18.28 & r    &  1.39 &  3.41 & $^{(1)}$   \\
GB6J001307+205335 & QSO & 3.53 & 00 13 11.10 +20 53 42.7 & 17.79 & g    &  1.11 &  4.02 & LBT06/18  \\
GB6J002121+155127 & QSO & 3.70 & 00 21 20.06 +15 51 25.8 & 19.59 & g    &  1.23 &  4.48 & $^{(1)}$   \\
GB6J003126+150729 & QSO & 4.29 & 00 31 26.80 +15 07 39.5 & 20.23 & g    &  1.36 &  3.47 & $^{(1)}$   \\
GB6J004348+342617 & QSO & 0.97 & 00 43 48.85 +34 26 26.1 & 19.06 & g    &  1.32 &  4.38 & $^{(10)}$  \\
GB6J010135+153845 & QSO & 1.50 & 01 01 36.09 +15 38 38.2 & 20.33 & i    &  1.47 &  4.11 & LBT11/17        \\
GB6J012126+034646 & QSO & 4.13 & 01 21 26.15 +03 47 06.7 & 18.77 & g    &  1.25 &  4.08 & LBT11/17        \\
GB6J012202+030951 & QSO & 4.00 & 01 22 01.91 +03 10 02.4 & 20.86 & g    &  1.12 & $<$3.88 & $^{(3)}$   \\
GB6J012921+310303 & QSO & 3.56 & 01 29 21.85 +31 02 58.5 & 19.26 & g    &  1.20 &  3.47 & $^{(1)}$   \\
GB6J024612+182334 & QSO & 3.59 & 02 46 11.82 +18 23 30.1 & 19.22 & g    &  1.15 &  4.52 & $^{(11)}$  \\
GB6J040738+001713 & QSO & 3.70 & 04 07 36.61 +00 17 26.3 & 19.60 & g    &  1.26 &  4.60 & TNG03/18        \\
GB6J042423+144230 & QSO & 3.55 & 04 24 23.49 +14 42 16.7 & 20.14 & g    &  1.26 &  4.80 & TNG03/18        \\
GB6J061110+721814 & $-$ & $-$  & 06 11 09.17 +72 18 15.6 & 19.78 & g    &  1.45 &  4.96 & LBT02/18        \\
GB6J064057+671228 & $-$ & $-$  & 06 40 58.19 +67 12 25.1 & 20.94 & r    & $>$1.76 &  3.46 & TNG03/18        \\
GB6J083548+182519 & QSO & 4.41 & 08 35 49.43 +18 25 20.1 & 20.93 & g    & $>$1.87 & $<$3.76 & $^{(1)}$   \\
GB6J083945+511206 & QSO & 4.40 & 08 39 46.22 +51 12 02.8 & 19.28 & g    &  1.98 &  4.06 & $^{(1,12)}$   \\
GB6J090631+693027 & QSO & 5.47 & 09 06 30.75 +69 30 30.8 & 20.54 & r    & $>$2.16 &  3.98 & $^{(5)}$   \\
GB6J091825+063722 & QSO & 4.22 & 09 18 24.38 +06 36 53.4 & 19.68 & g    &  1.40 &  4.09 & $^{(1,12)}$   \\
GB6J100724+580201 & QSO & 3.77 & 10 07 24.88 +58 02 03.5 & 17.59 & g    &  1.23 &  4.05 & $^{(1)}$   \\
GB6J102623+254255 & QSO & 5.28 & 10 26 23.62 +25 42 59.4 & 20.06 & r    &  1.77 &  4.28 & $^{(1,12)}$   \\
GB6J132512+112338 & QSO & 4.42 & 13 25 12.49 +11 23 29.8 & 19.45 & g    &  2.13 &  4.13 & $^{(1,12)}$   \\
GB6J134811+193520 & QSO & 4.40 & 13 48 11.26 +19 35 23.5 & 20.64 & g    &  1.53 &  3.48 & $^{(1)}$   \\
GB6J141212+062408 & QSO & 4.47 & 14 12 09.97 +06 24 06.8 & 20.19 & g    &  1.29 &  2.99 & $^{(1,12)}$   \\
GB6J143023+420450 & QSO & 4.72 & 14 30 23.74 +42 04 36.5 & 19.79 & r    &  1.27 & $<$2.88 & $^{(1)}$   \\
GB6J143533+543605 & QSO & 3.81 & 14 35 33.78 +54 35 59.3 & 20.20 & g    &  1.13 &  3.45 & $^{(1)}$   \\
GB6J151002+570256 & QSO & 4.31 & 15 10 02.93 +57 02 43.4 & 20.52 & g    &  1.21 &  4.48 & $^{(1,12)}$   \\
GB6J154929+170853 & BL  & 1.2: & 15 49 29.27 +17 08 28.0 & 18.77 & g    & $>$4.03 &  3.80 & $^{(1)}$   \\
GB6J155930+030444 & QSO & 3.89 & 15 59 30.98 +03 04 48.3 & 20.12 & g    &  1.29 &  4.25 & $^{(1)}$   \\
GB6J161216+470311 & QSO & 4.36 & 16 12 16.76 +47 02 53.6 & 20.53 & g    &  1.28 &  3.80 & $^{(1)}$   \\
GB6J162956+095959 & QSO & 5.00 & 16 29 57.28 +10 00 23.5 & 20.77 & r    &  1.17 & $<$3.87 & LBT05/18        \\
GB6J164327+410359 & QSO & 3.86 & 16 43 26.24 +41 03 43.5 & 20.05 & g    &  1.11 &  4.19 & $^{(1)}$   \\
GB6J164856+460341 & QSO & 5.36 & 16 48 54.53 +46 03 27.4 & 20.31 & r    &  1.48 &  3.86 & LBT01/18        \\
GB6J171103+383016 & QSO & 4.00 & 17 11 05.54 +38 30 04.3 & 20.53 & g    &  1.12 &  4.25 & LBT01/18        \\
GB6J205332+010307 & QSO & 3.59 & 20 53 31.73 +01 03 42.2 & 19.72 & g    &  1.18 &  4.49 & TNG05/18        \\
GB6J223927+225959 & QSO & 2.93 & 22 39 27.69 +23 00 18.1 & 18.17 & g    &  1.92 &  3.97 & $^{(1)}$   \\
GB6J231449+020146 & QSO & 4.11 & 23 14 48.71 +02 01 51.1 & 19.64 & g    &  1.57 &  3.53 & $^{(8)}$   \\
GB6J235758+140205 & QSO & 4.35 & 23 57 58.56 +14 02 01.9 & 20.40 & g    &  1.11 & $<$3.22 & LBT07/18   \\
\hline
\end{tabular}

{\bf column 1}: name; {\bf column 2}: CLASS (QSO=broad emission line AGN; BL=BL Lac object); {\bf column 3}: redshift (``:''=tentative redshift); {\bf column 4}: radio (8.4~GHz, VLA-A array) position (J2000.0) {\bf column 5}: PS1 magnitude in the reddest filter of the dropout
(i.e. {\it r}-filter for {\it g}-dropout sources, {\it i}-filter for {\it r}-dropout sources, {\it z}-filter for {\it i}-dropout sources;
{\bf column 6}: type of dropout; {\bf column 7}: dropout value; {\bf column 8}: {\it z}-W2; {\bf column 9}: reference for the optical spectrum (or to the spectroscopic identification): (1)=SDSS DR12, (2)=\citet{Stern2000}, (3)=\citet{Sowards-Emmerd2002}, (4)=\citet{Amirkhanyan2006}, (5)=\citet{Romani2004}, (6)=\citet{Healey2008}, (7)=\citet{Hook1998}, (8)=\citet{Hook2002}; (9)=\citet{Hewett2010a}; 
(10)=\citet{Ackermann2011}; (11)=BZCAT; (12)=\citet{Sbarrato2013a}. Please note that some of the 
objects with an SDSS spectrum have been already published in the literature.
\end{table*}
%--------------------------------------------------------------------------------------------

\section{The spectroscopic observations}
During 2017-2018 we have carried out dedicated spectroscopic observations 
of the high-z blazar candidates of CLASS using the Large Binocular
Telescope (LBT) and the Telescopio Nazionale Galileo (TNG).
In the following, we give some 
detail on these observing runs. The new spectroscopic identifications collected
at these telescopes are reported in Tab.~\ref{dropout_sample}

{\it Large Binocular Telescope (LBT)}. LBT has a binocular design with two
identical 8.4 meter telescopes and it is located on Mount Graham,  Arizona (US).
For our project, we used LBT coupled to the Multi-Object Double Spectrograph
(MODS , Pogge et al. 2010). 
The observations were taken during the period November 2017-May 2018 
with the red grating (G670L, 5000-10000\AA)
and using slit widths of 1-1.2$\arcsec$.
Data reduction was performed at the Italian LBT Spectroscopic Reduction
Center through scripts optimized for LBT data. Each spectral image was independently 
bias subtracted and flat-field corrected. Sky subtraction was done on 2D
extracted, wavelength calibrated spectra with a fit. Wavelength calibration
was obtained from spectra of arc lamps (rms=0.08$\AA$ on MODS1 and 0.1$\AA$ on
MODS2), while flux calibration was achieved from observations of a
spectro-photometric standard. 
%(see more details in Magrini et
%al 2012). 
%In Fig.~\ref{spectra} we report the optical spectra of the high-z AGN with an SDSS spectrum plus
%the object observed at LBT.

{\it Telescopio Nazionale Galileo (TNG)}.
The TNG telescope is a 3.58-meter telescope located in the island of
San Miguel de La Palma (Spain). We carried out the observations in March and
in May 2018 using
DOLORES (Device Optimized for the LOw RESolution) which is a focal reducer
instrument installed at the Nasmyth B focus of the telescope. We used
the LR-R/LR-B  grisms and a long-slit with a width between 1$\arcsec$ and
1.5$\arcsec$, depending to the actual seeing conditions, oriented along
the parallactic angle.
For the data reduction we have used the IRAF longslit package. The
spectra have been wavelength calibrated using Ar, Ne+Hg, Kr reference spectra. 
The flux calibration was obtained by observing a spectrophotometric standard. 

In total, we observed all the 12 sources without
a spectroscopic redshift. In 10 cases the
data were good enough to obtain a firm classification and redshift while for
2 sources the signal-to-noise is too low and they should be considered still unidentified.

In nearly all cases the sources are confirmed high-z AGN (z$>$3) although not 
all  have a redshift above the adopted
threshold (z=4). In one case (GB6J010135+153845), instead, the redshift is significantly lower
than expected (z=1.5). This was the only {\it i}-dropout candidate for which we predicted a very
high redshift (z$\sim$5.5-6.5). The optical spectrum, instead, shows a broad emission
line at $\lambda$=7075\AA\ that can be identified as the MgII$\lambda$2798\AA, leading to a redshift of 1.53.
The continuum emission does not show any obvious dropout between the {\it i} and {\it z} band contrary to the
value reported by the PS1 catalogue ({\it i$-$z}=1.47$\pm$0.07 mag). Very likely this a fake 
dropout due to some errors in the PS1 magnitudes.

Notably, with these observations {\it we have found two new z$\geq$5 AGN thus doubling the number
of AGN in this redshift bin present in the CLASS sample}. As described in Section~5, both 
objects have a radio spectrum that confirms their blazar nature. 
Therefore, they constitute a significant
addition with respect to the total number of blazars at z above 5 discovered so far (4 in total, see also Belladitta et al. in prep).
One source (GB6J164856+460341), in particular, has a redshift of 5.38 and it represents {\it the second most distant blazar discovered so far} (the highest
being at z=5.47, GB6J090631+693027, \citealt{Romani2004}). 

We have not found, instead, high redshift 
featureless blazars (i.e. BL Lac objects) even though our selection is potentially 
sensitive also to these objects.
This lack of high-z BL Lac is likely due to their peculiar cosmological 
evolution (very weak or even negative, see e.g. \citealt{Ajello2014} and references therein) 
that makes their detection at high redshifts very unlikely. In the following, with the
term ``blazar'' we will always indicate the QSO-like blazars (often called flat-spectrum radio 
quasar, FSRQ) due to the lack of featureless blazars at high redshifts.

In total, considering both the new identifications and those from the literature or from the SDSS,
we have 20 QSO with z$\geq$4 among the 37 candidates selected with the dropout method. Only
two objects (5\%) are still unidentified. 
Six sources are in common with the sample
of high-z blazar candidates selected by 
\citet{Sbarrato2013a}.

All the optical spectra of the z$\geq$4 objects observed at LBT and TNG or with an optical spectrum from SDSS
are reported in Fig.~\ref{spectra}. A detailed analysis of these data is in progress. In particular,
we are measuring the line widths and luminosities of the most important emission lines, like Ly$\alpha$, C$\small{IV}\lambda$1549\AA\ and C$\small{III]}\lambda1909\AA$ (when observed) with the final goal of obtaining a reliable estimate of the mass of central SMBH and of the accretion rate.

The redshift distribution of all the z$\geq$4 objects in CLASS is reported in Fig.~\ref{z_dist}.

%--------------------------------------------------------------------------------------------
\begin{table*}
\caption{Other z$>$4 QSO present in the CLASS survey}
\label{other}
\begin{tabular}{c c c c c r r r r}
\hline\hline
 name             & redshift & coordinates       & mag   & dropout & drop value & z$-$W2 & Ref & b$^{II}$\\ 
                  &          &   (J2000.0)       & (mag) &         & (mag)      &  (mag) &        & (deg) \\
\hline
GB6J025758+433837 &  4.07 & 02 57 59.08 +43 38 37.7 & 19.86 & g    &     1.04 &  3.25 & $^{(4)}$   & -13.5101 \\
GB6J102107+220904 &  4.26 & 10 21 07.58 +22 09 21.6 & 21.46 & g    & $>$   1.34 & $<$4.16 & $^{(1)}$   &  55.6110 \\
{\bf GB6J153533+025419} &  4.39 & 15 35 33.88 +02 54 23.4 & 20.72 & g    &     0.79 &  3.33 & $^{(1)}$   &  43.9170 \\
GB6J160608+312504 &  4.56 & 16 06 08.52 +31 24 46.5 & 21.32 & r    & $>$   1.38 &  5.57 & $^{(6)}$   &  47.6632 \\
GB6J171521+214547 &  4.01 & 17 15 21.25 +21 45 31.7 & 21.51 & g    & $>$   1.29 &  - & $^{(7)}$   &  30.3650 \\
GB6J195135+013442 &  4.11 & 19 51 36.02 +01 34 42.7 & 20.56 & g    &     1.16 &  3.65 & $^{(6)}$   & -12.6097 \\
\hline
\end{tabular}

This table includes other z$>$4 QSO present in the CLASS survey that have not been selected mostly
because they have a low Galactic latitude or because they have magnitude below the adopted limit.
Only one object (in bold face) is missed by the selection because of a low dropout value. This should be considered 
as part of the CLASS sample of high-z blazar candidates since it falls in the sky area considered for the selection
and it has a magnitude brighter than the limit.
Table caption as in the previous table except for the last column that contains the Galactic latitude. 
\label{add}
\end{table*}
%%----------------------------------------------------------------------------------
%%---------------------------------------------------------------------------------------

   \begin{figure*}
   \centering
   \includegraphics[width=4.3cm, angle=-90]{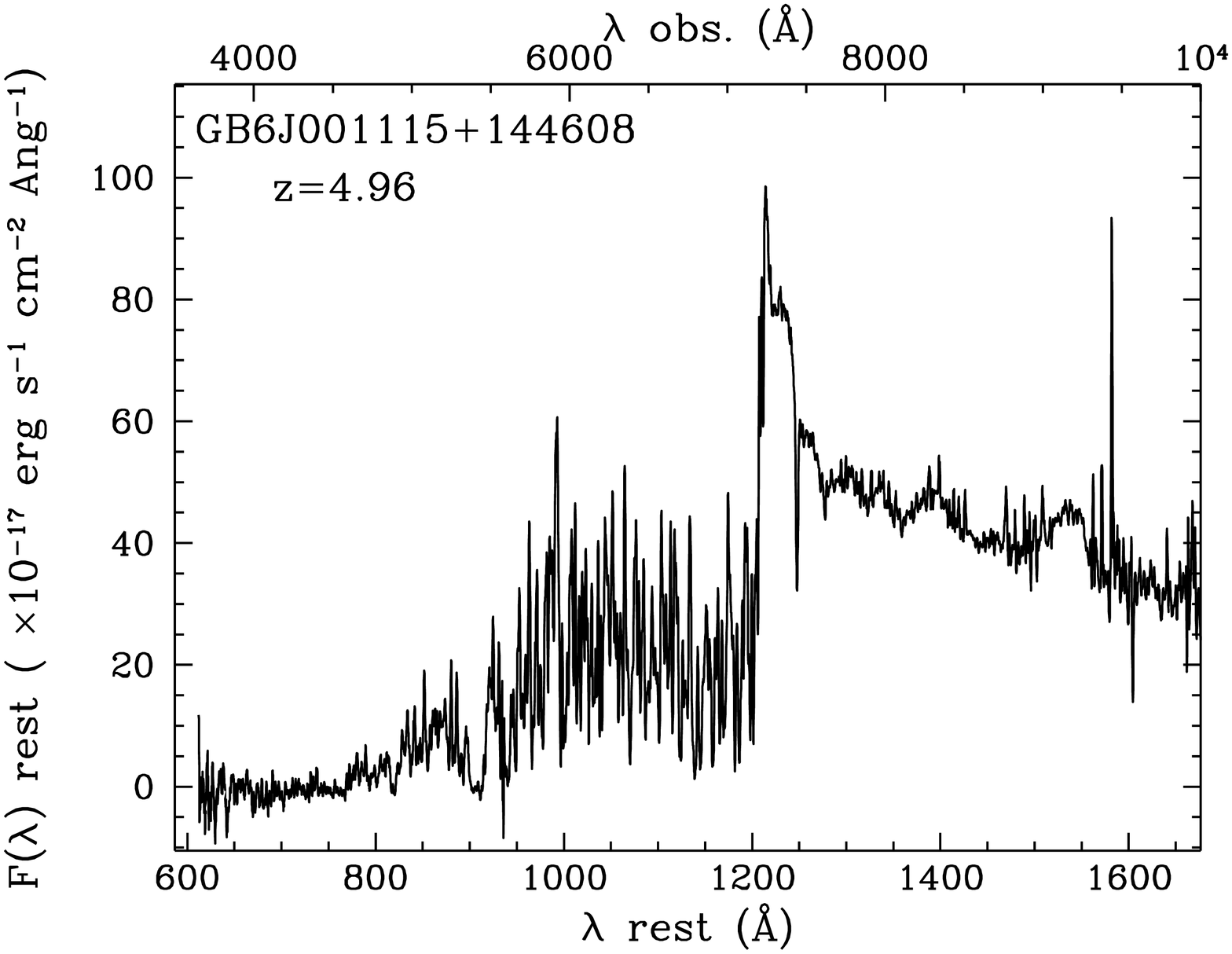}
    \includegraphics[width=4.3cm, angle=-90]{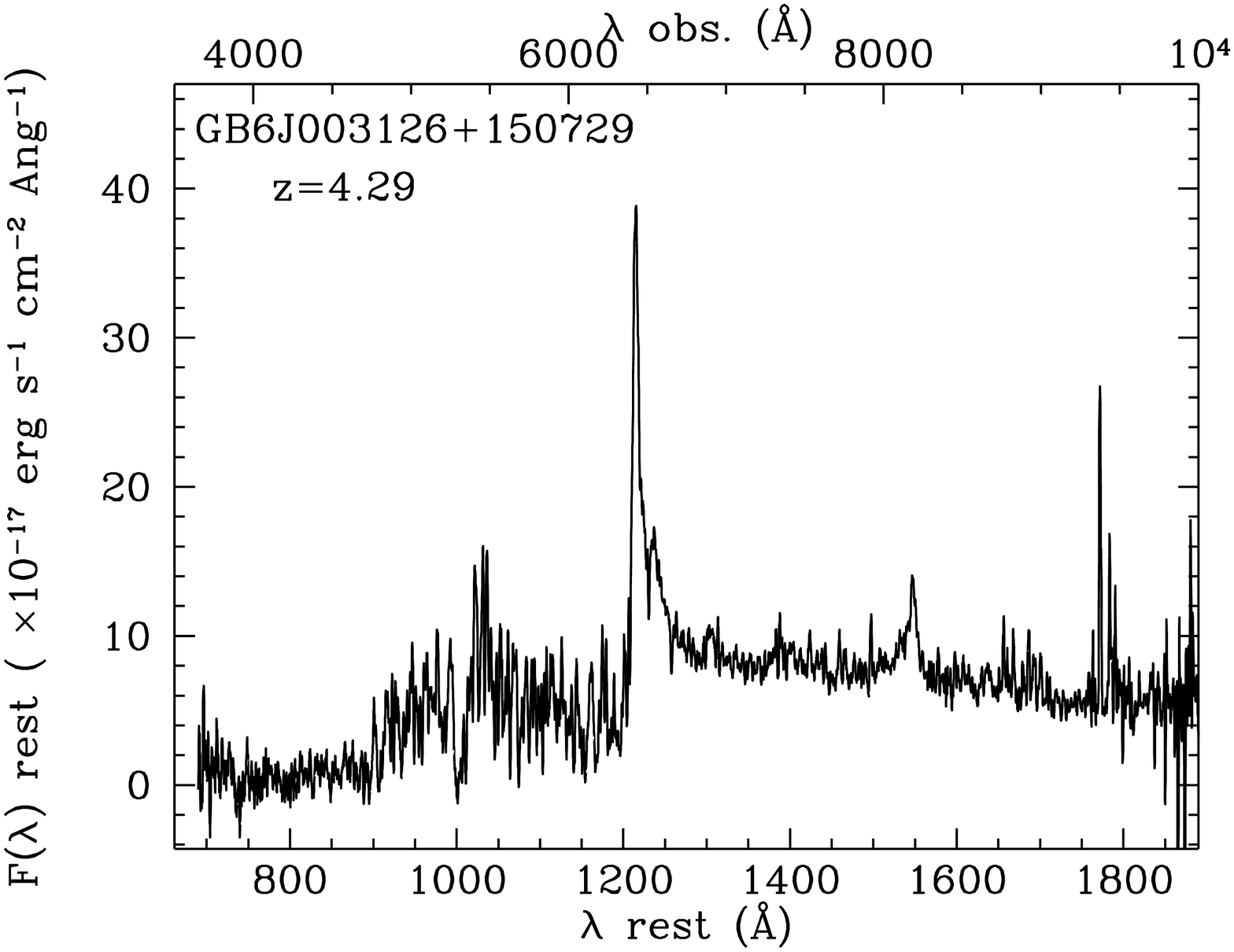} 
   \includegraphics[width=4.3cm, angle=-90]{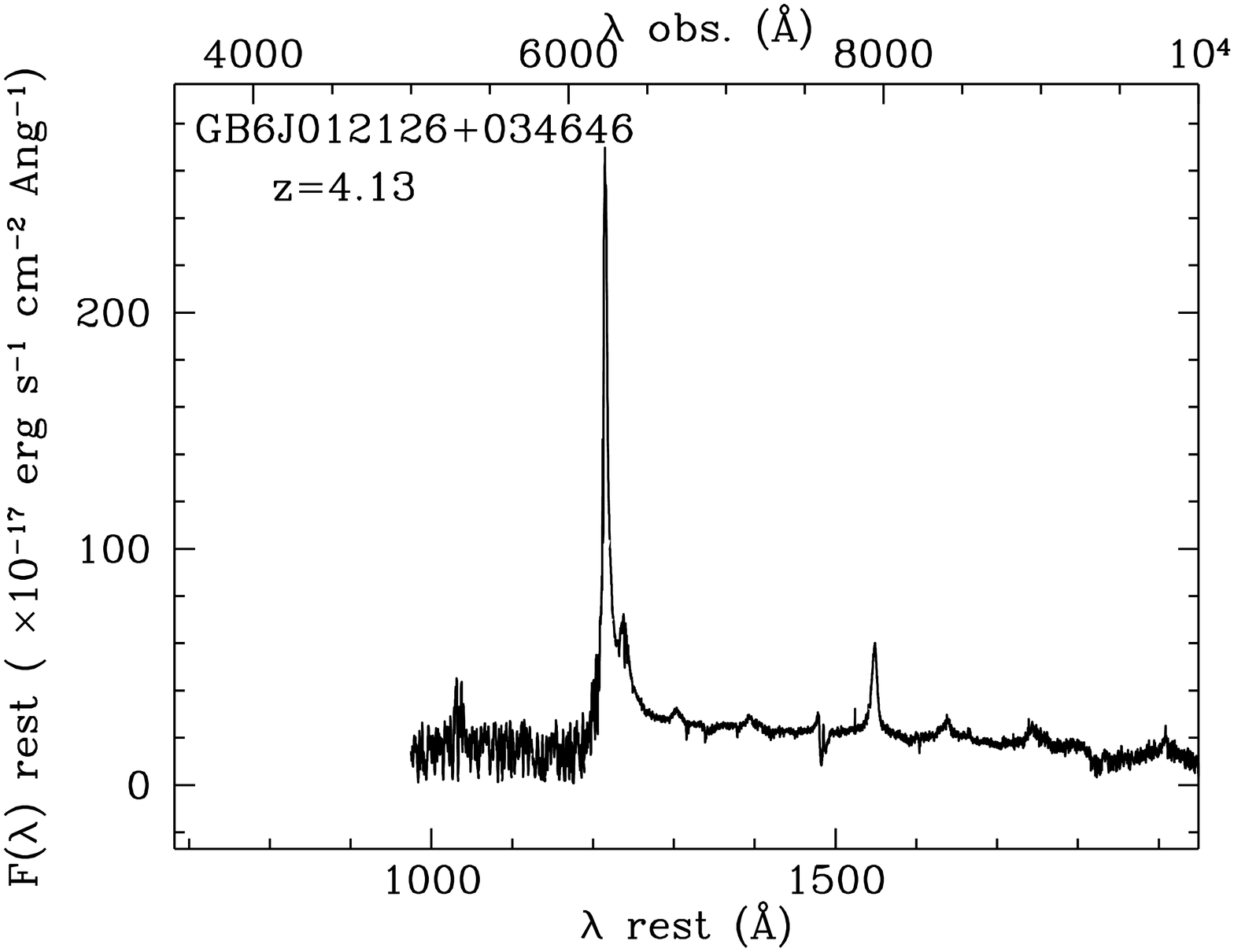}
   \includegraphics[width=4.3cm, angle=-90]{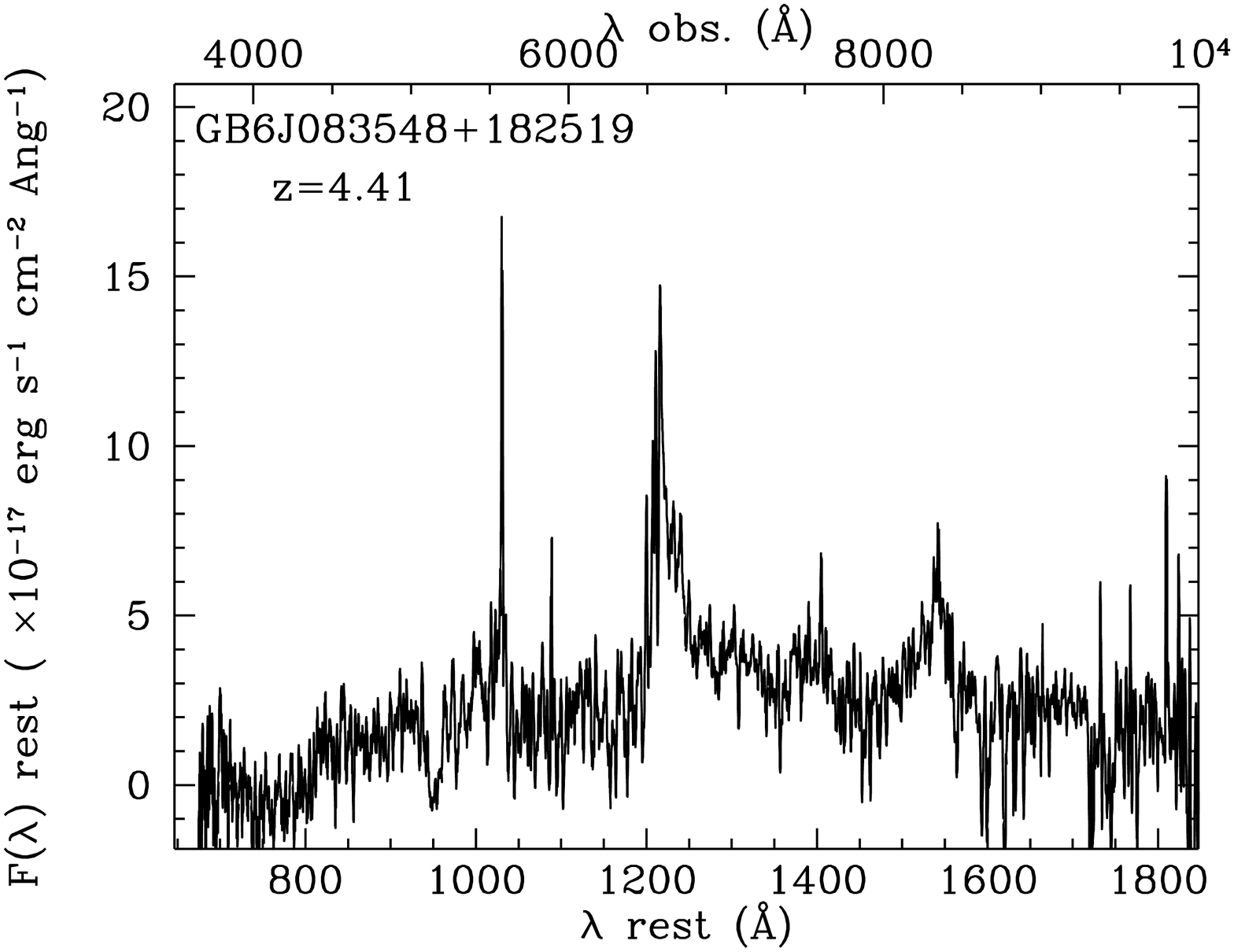}
   \includegraphics[width=4.3cm, angle=-90]{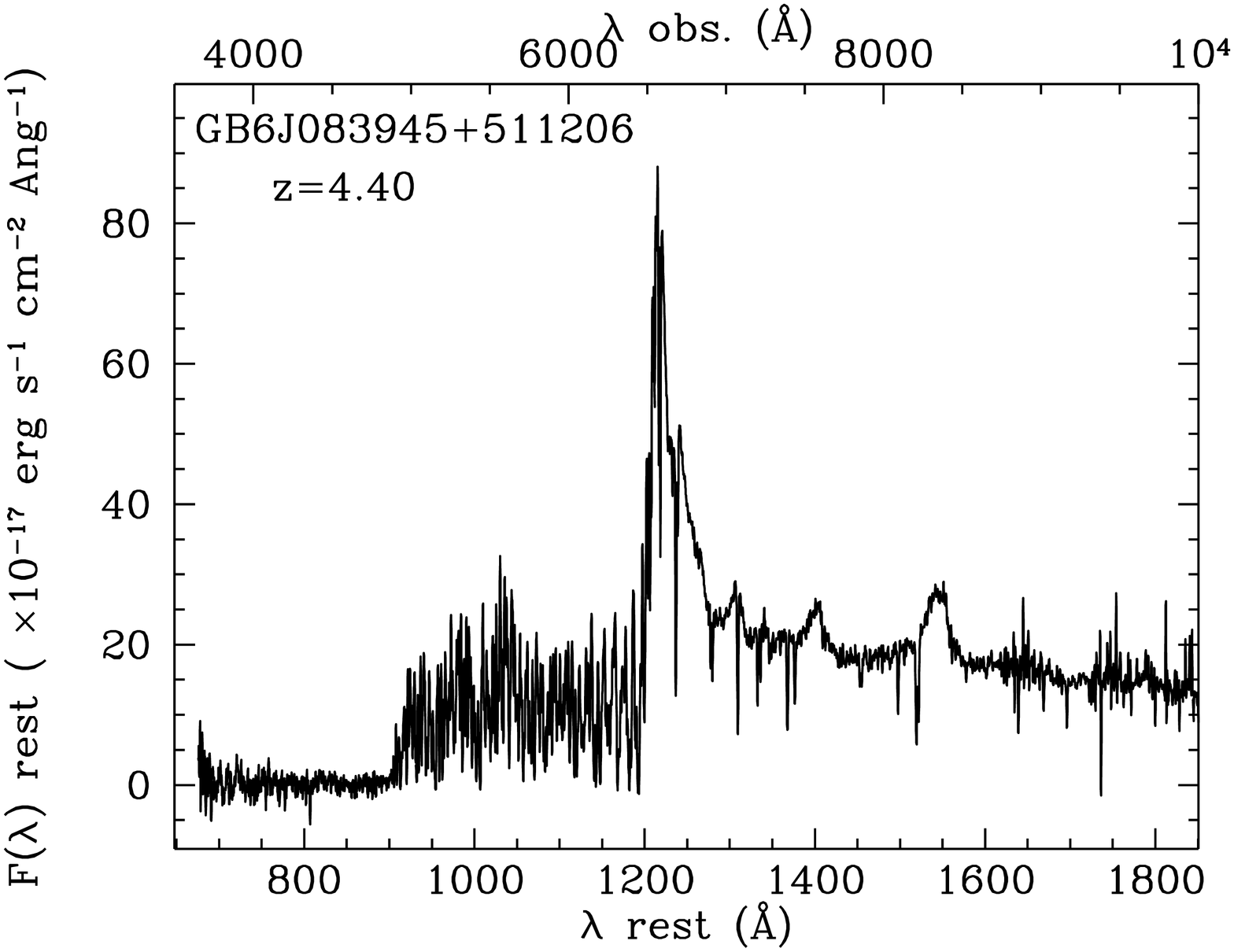}
\includegraphics[width=4.3cm, angle=-90]{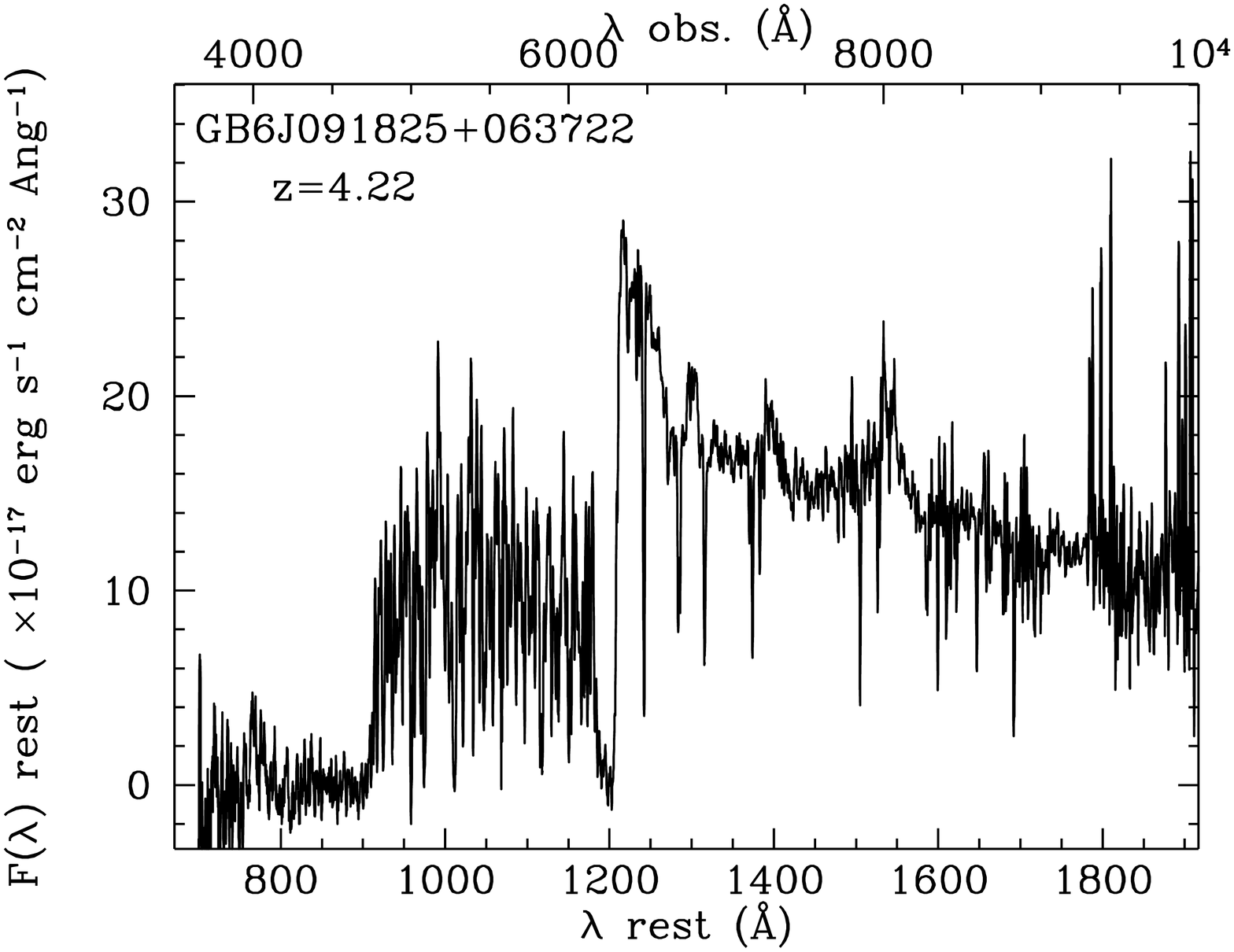}
\includegraphics[width=4.3cm, angle=-90]{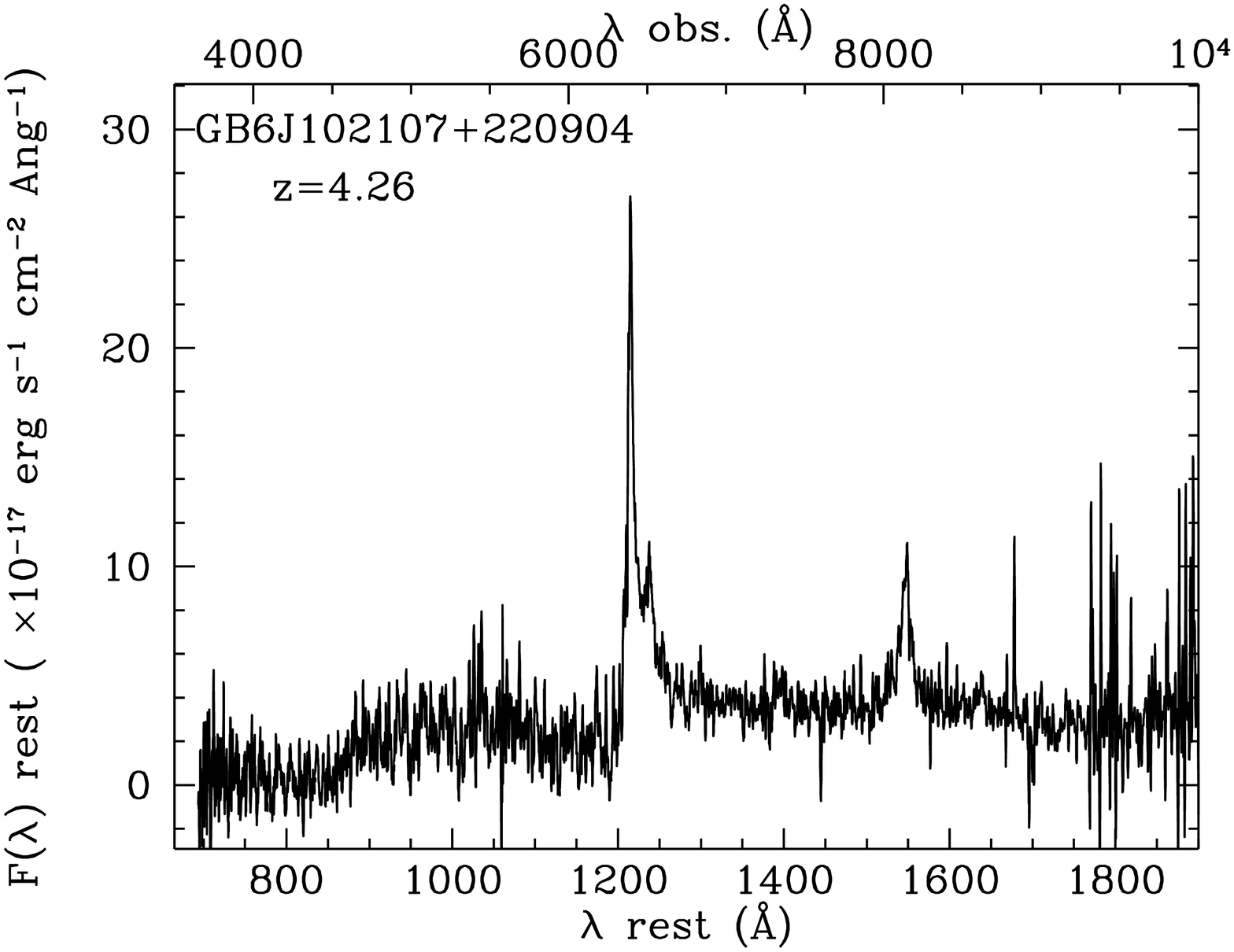}
\includegraphics[width=4.3cm, angle=-90]{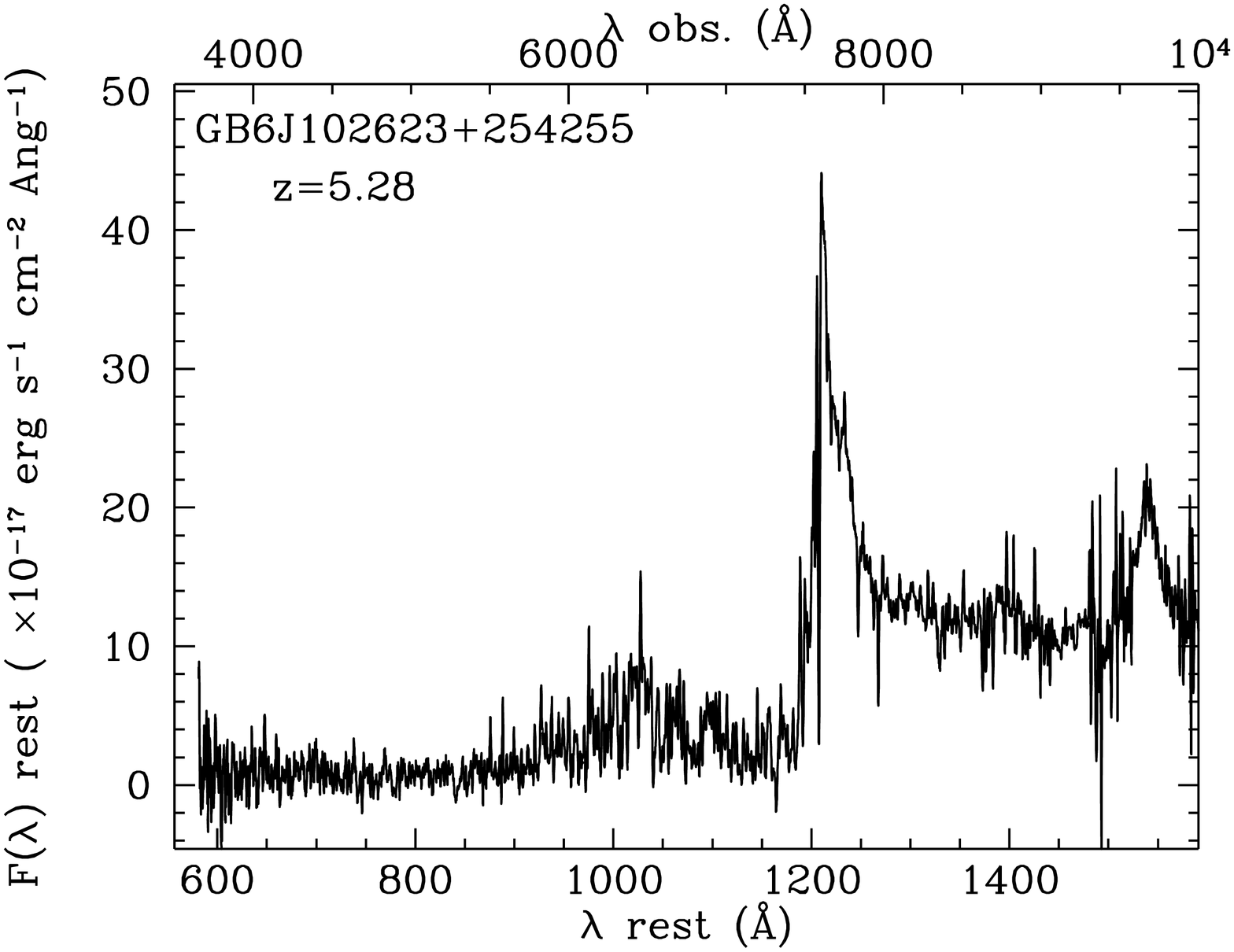}
\includegraphics[width=4.3cm, angle=-90]{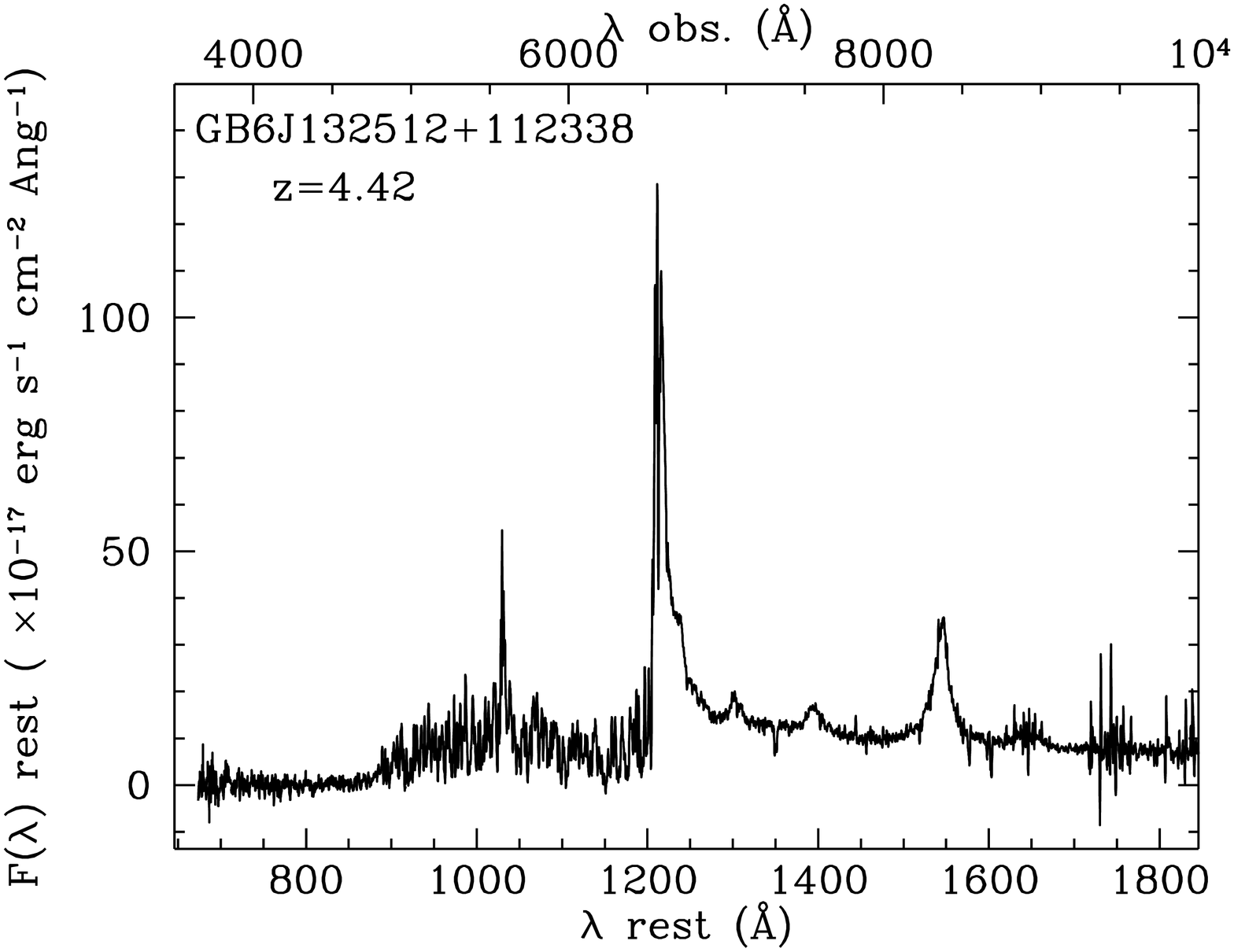}
\includegraphics[width=4.3cm, angle=-90]{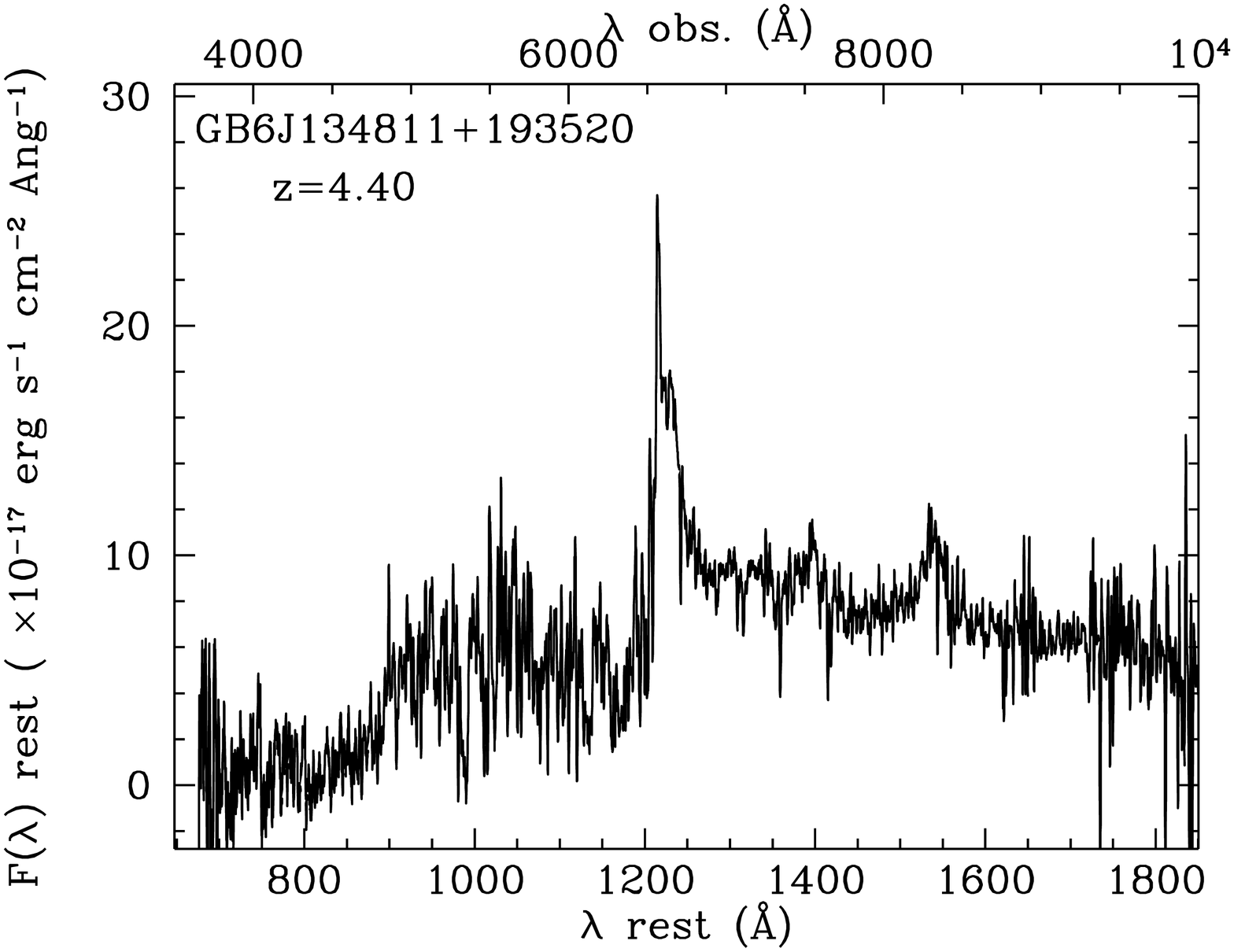}
\includegraphics[width=4.3cm, angle=-90]{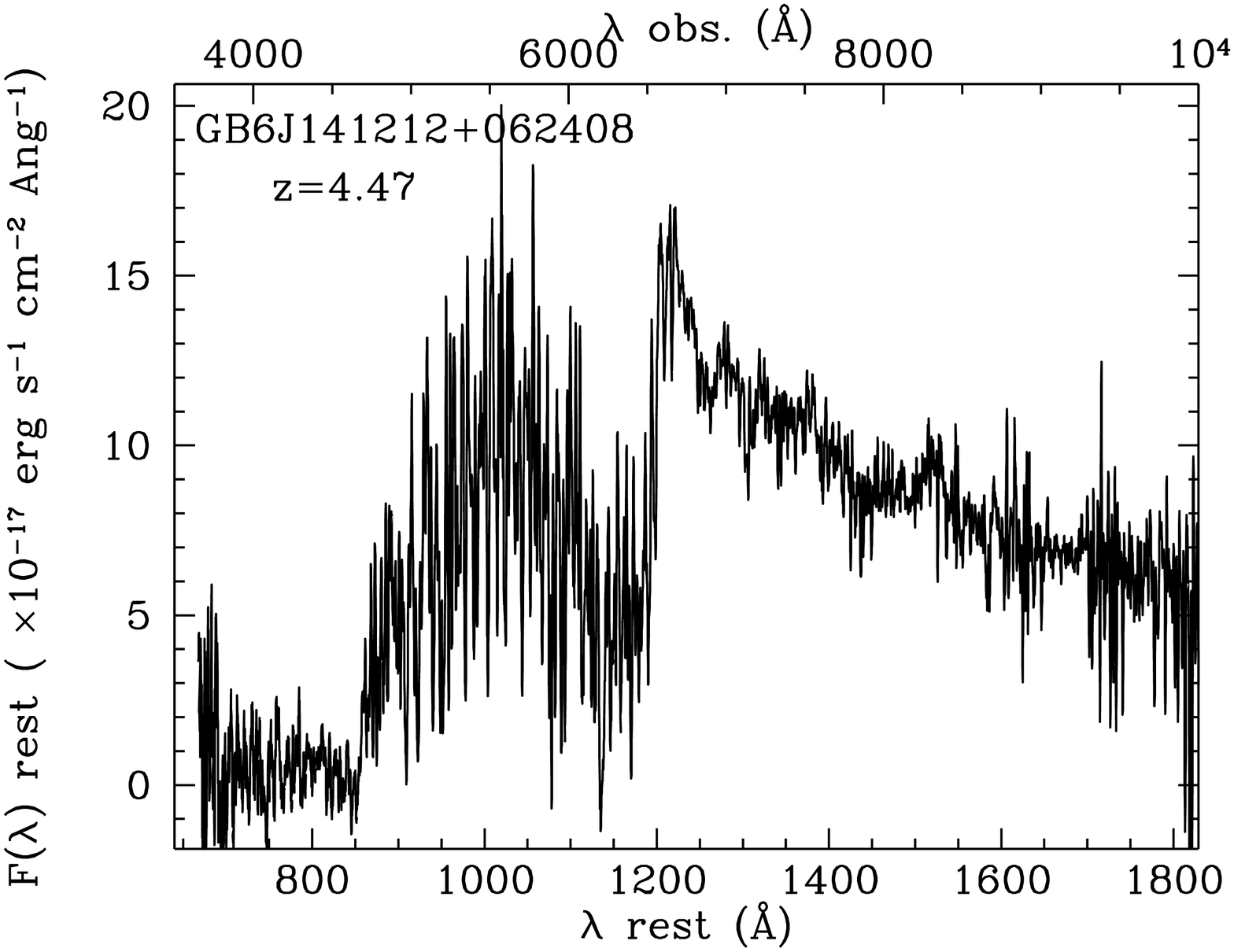}
\includegraphics[width=4.3cm, angle=-90]{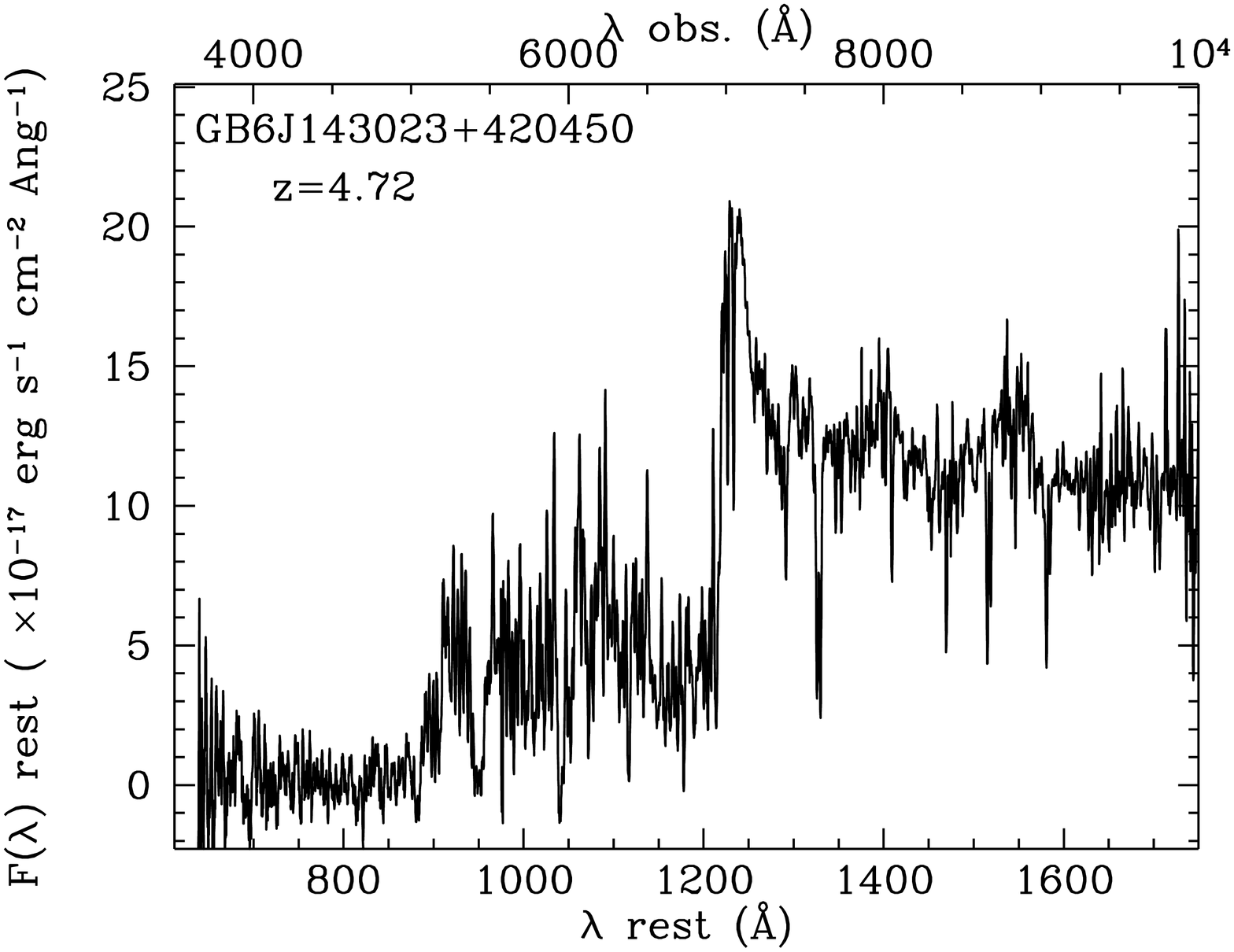}
   \caption{(continued)}
\end{figure*}
\addtocounter{figure}{-1}
   \begin{figure*}
     \centering
    \includegraphics[width=4.3cm, angle=-90]{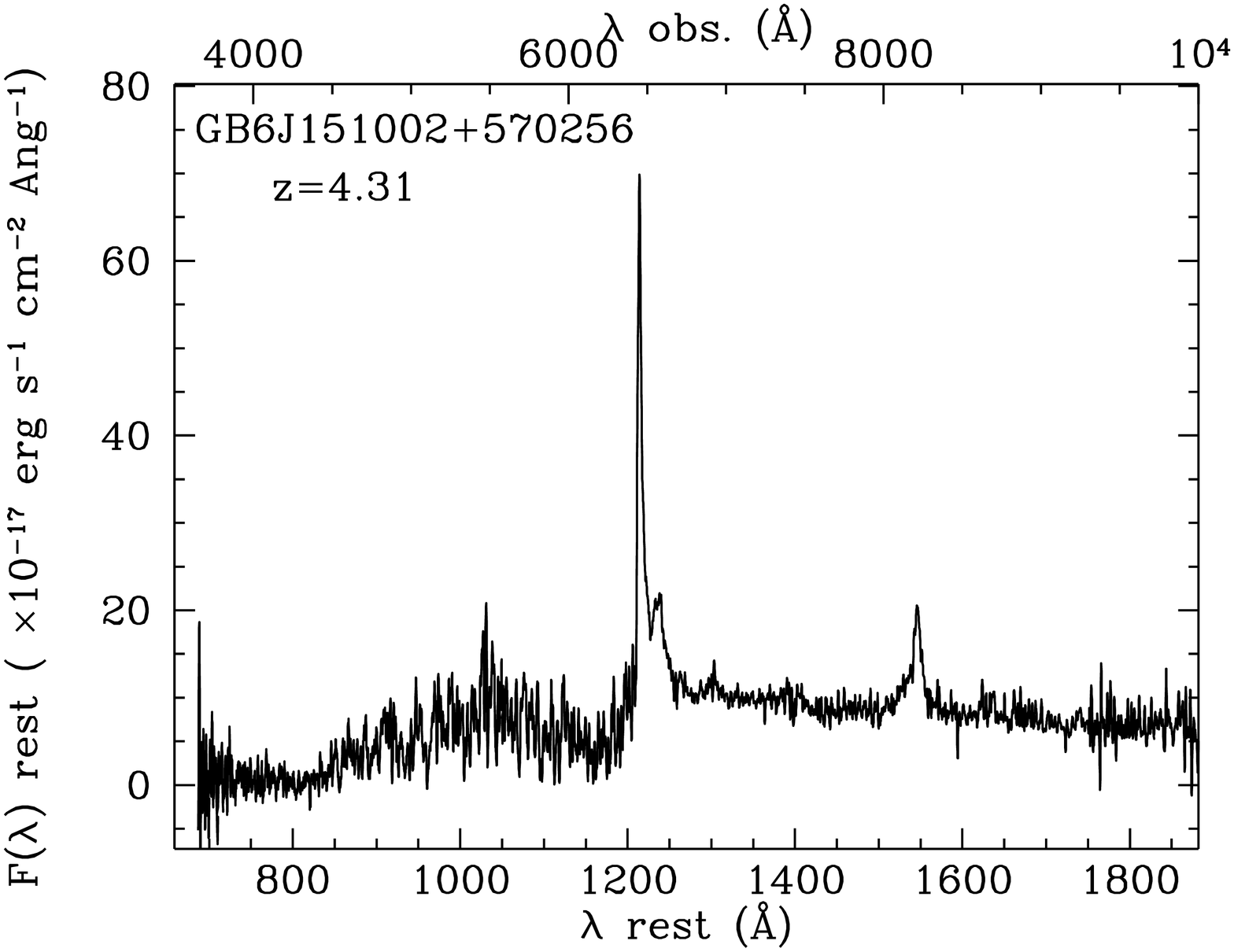}
    \includegraphics[width=4.3cm, angle=-90]{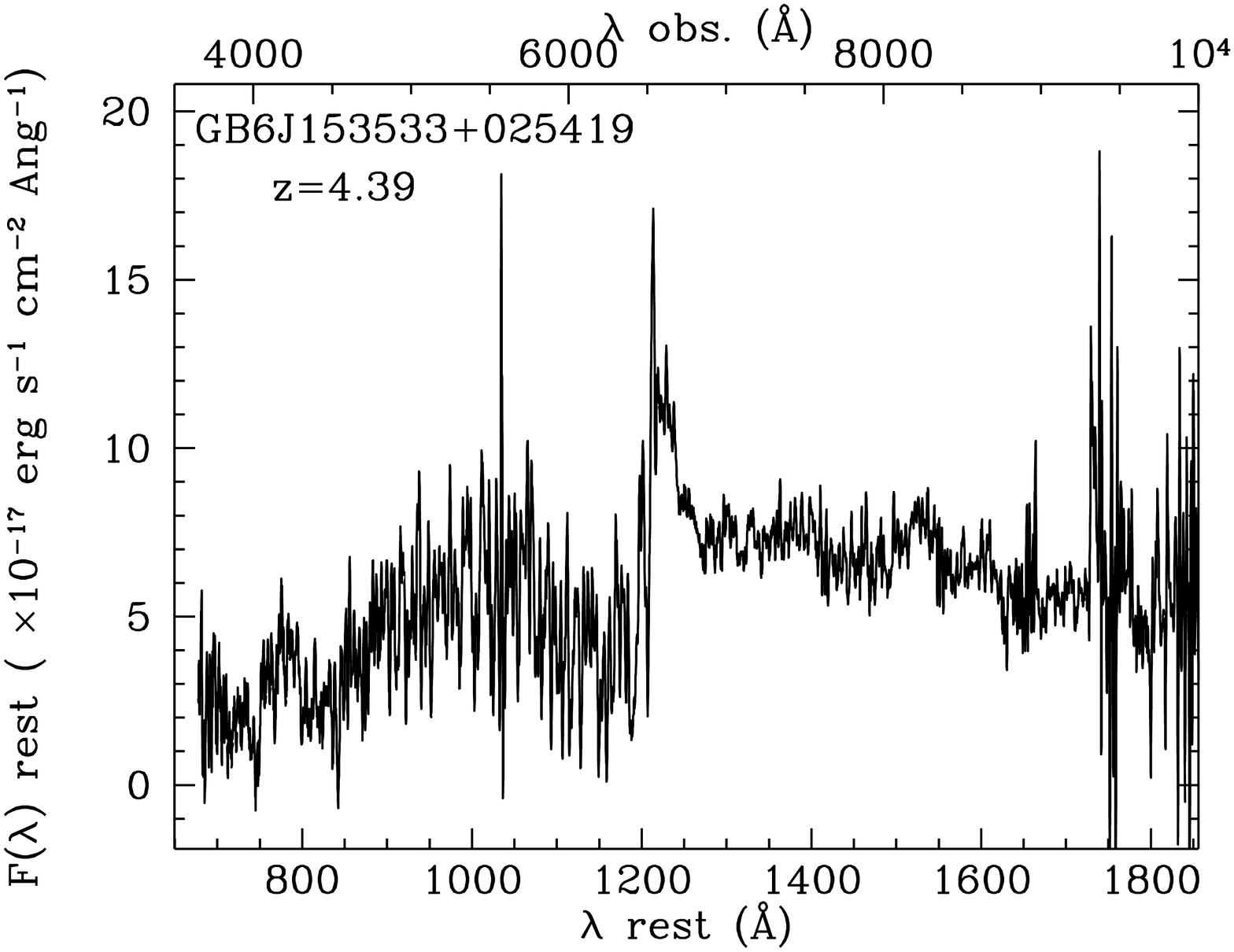}
    \includegraphics[width=4.3cm, angle=-90]{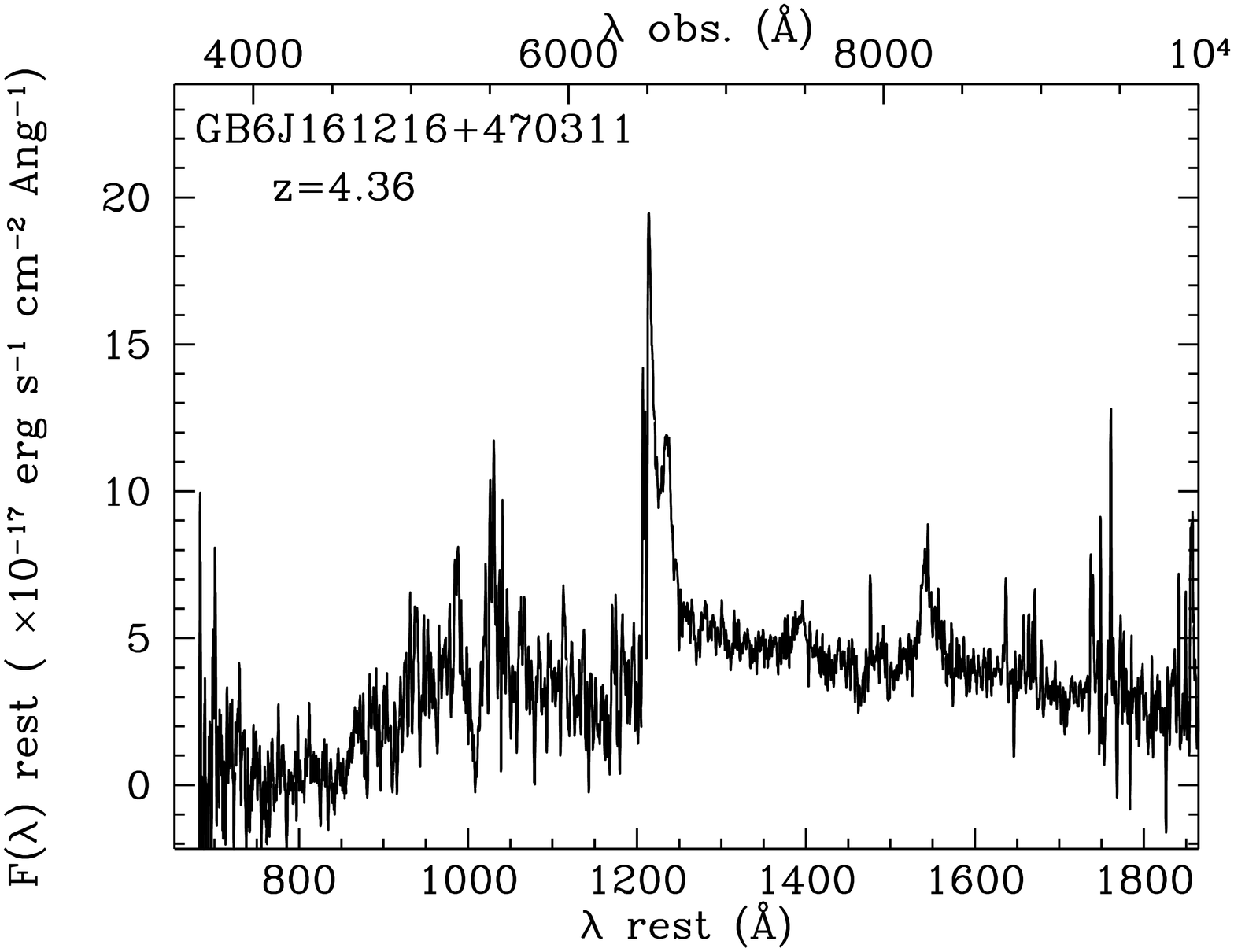}
    \includegraphics[width=4.3cm, angle=-90]{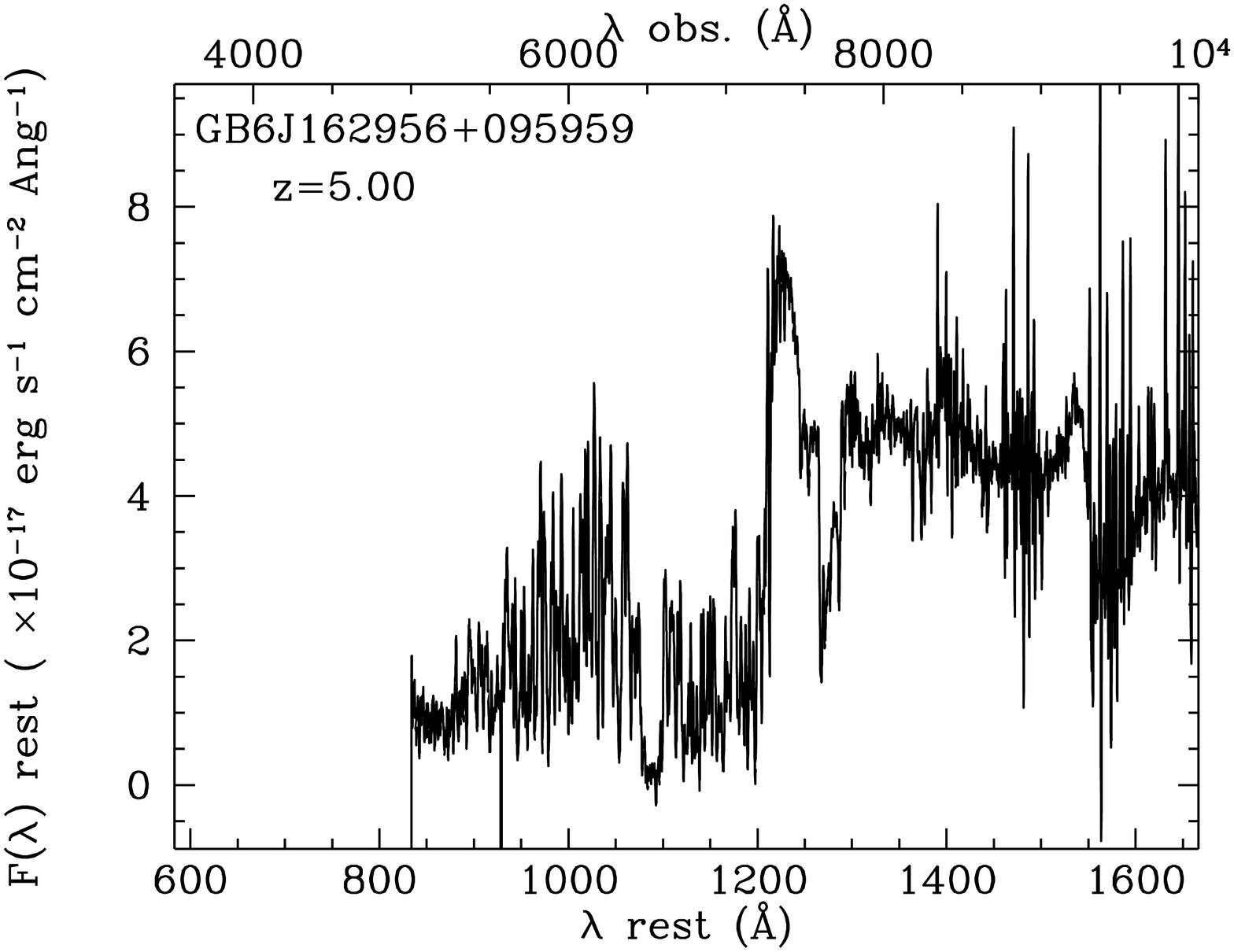}
   \includegraphics[width=4.3cm, angle=-90]{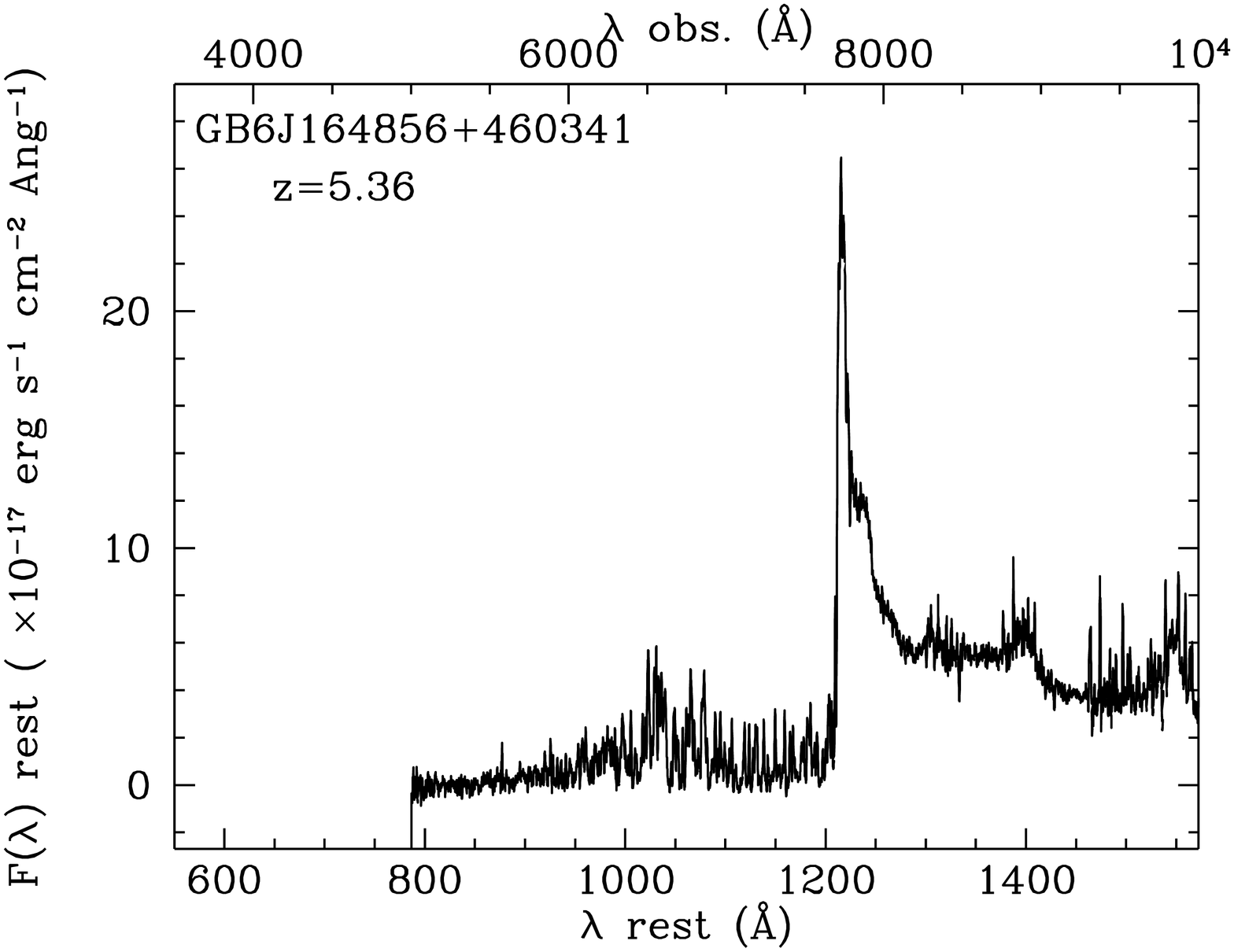}
    \includegraphics[width=4.3cm, angle=-90]{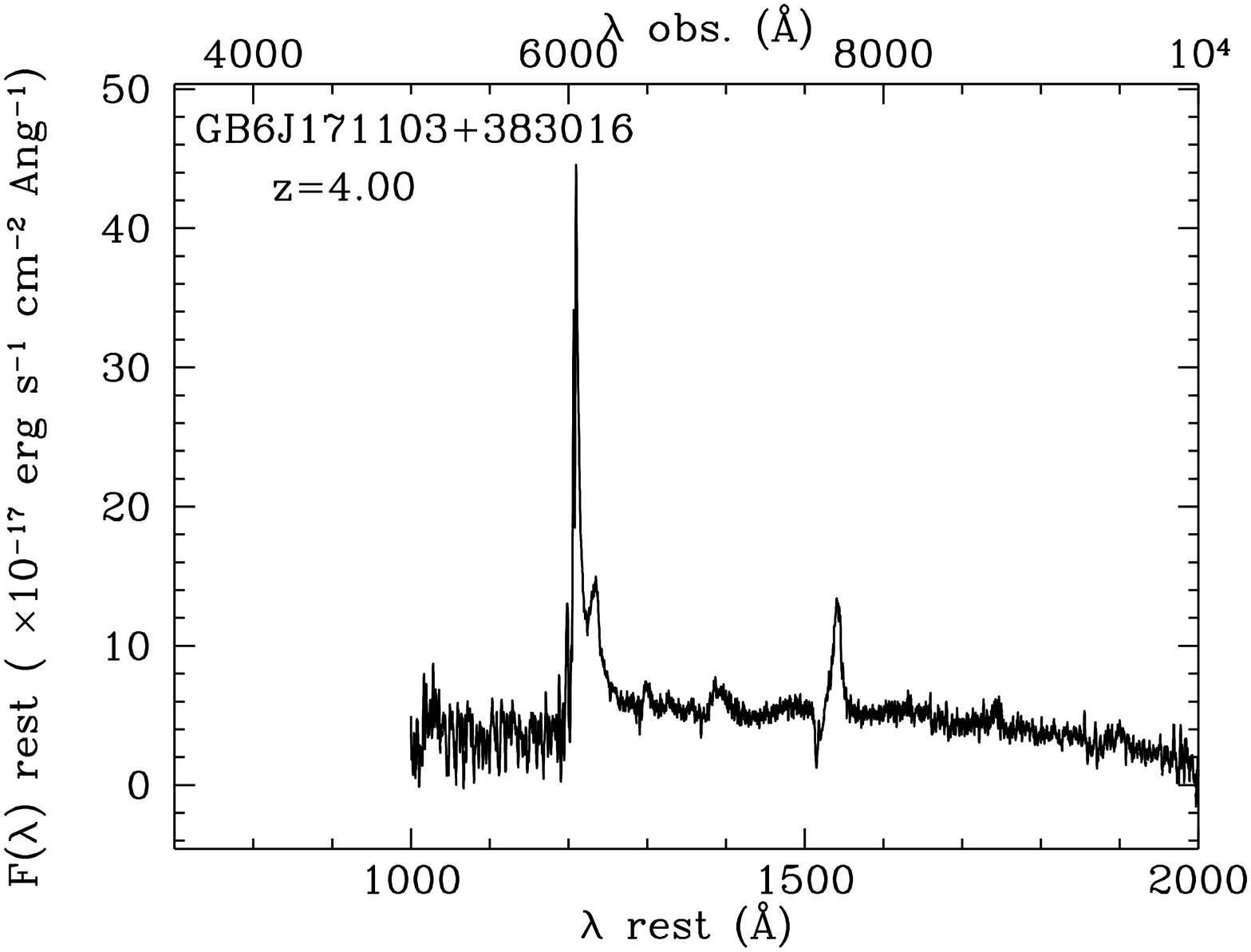}
    \includegraphics[width=4.3cm, angle=-90]{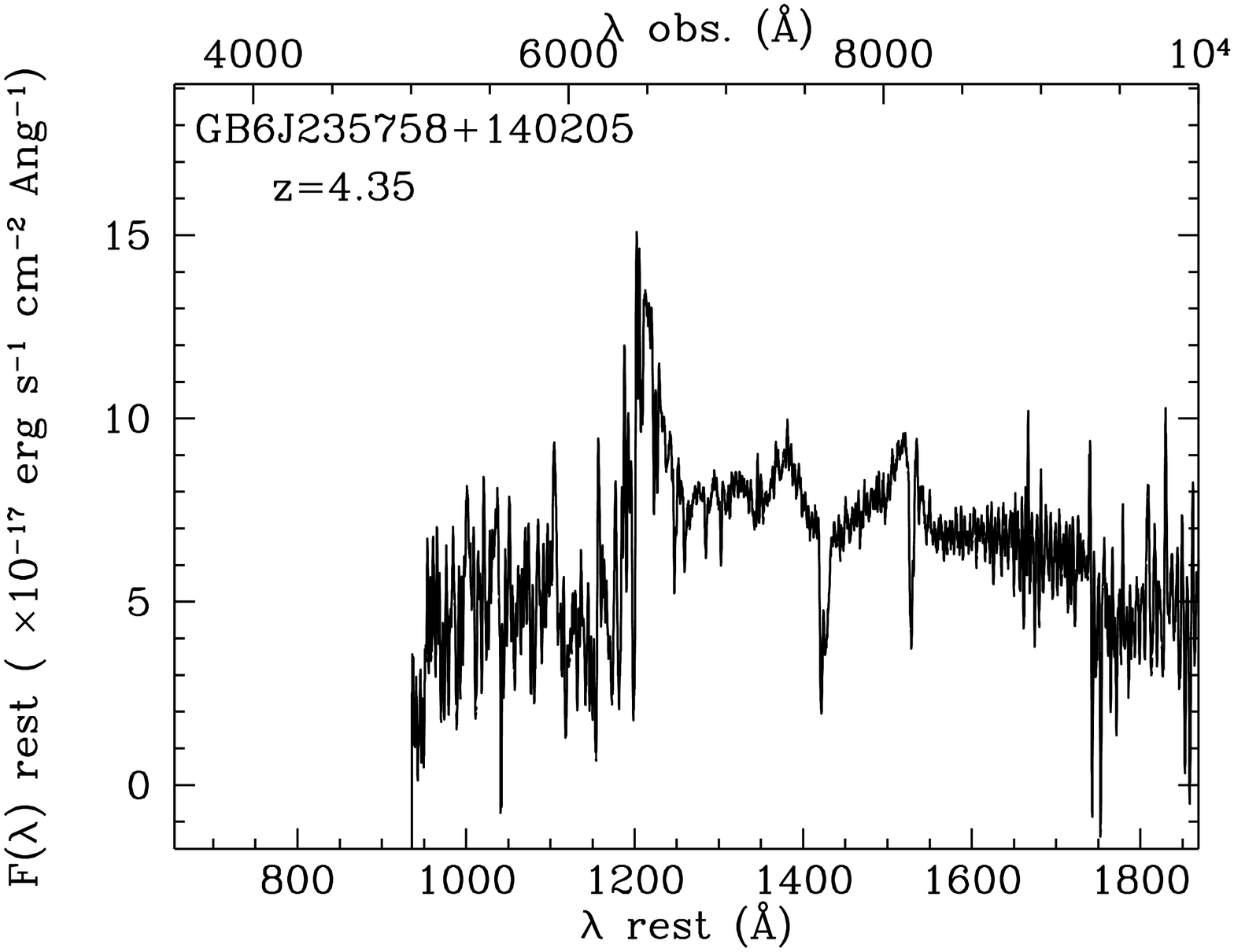}
 \caption{Optical spectra (either from SDSS or taken by our own observations) of 19 out of the 26 z$\geq$4 QSO 
present in the CLASS survey  (see text for details)}
 \label{spectra}
\end{figure*}

%--------------------------------------------------------------------------------------
  \begin{figure}
    \centering
    \includegraphics[width=6cm, angle=0]{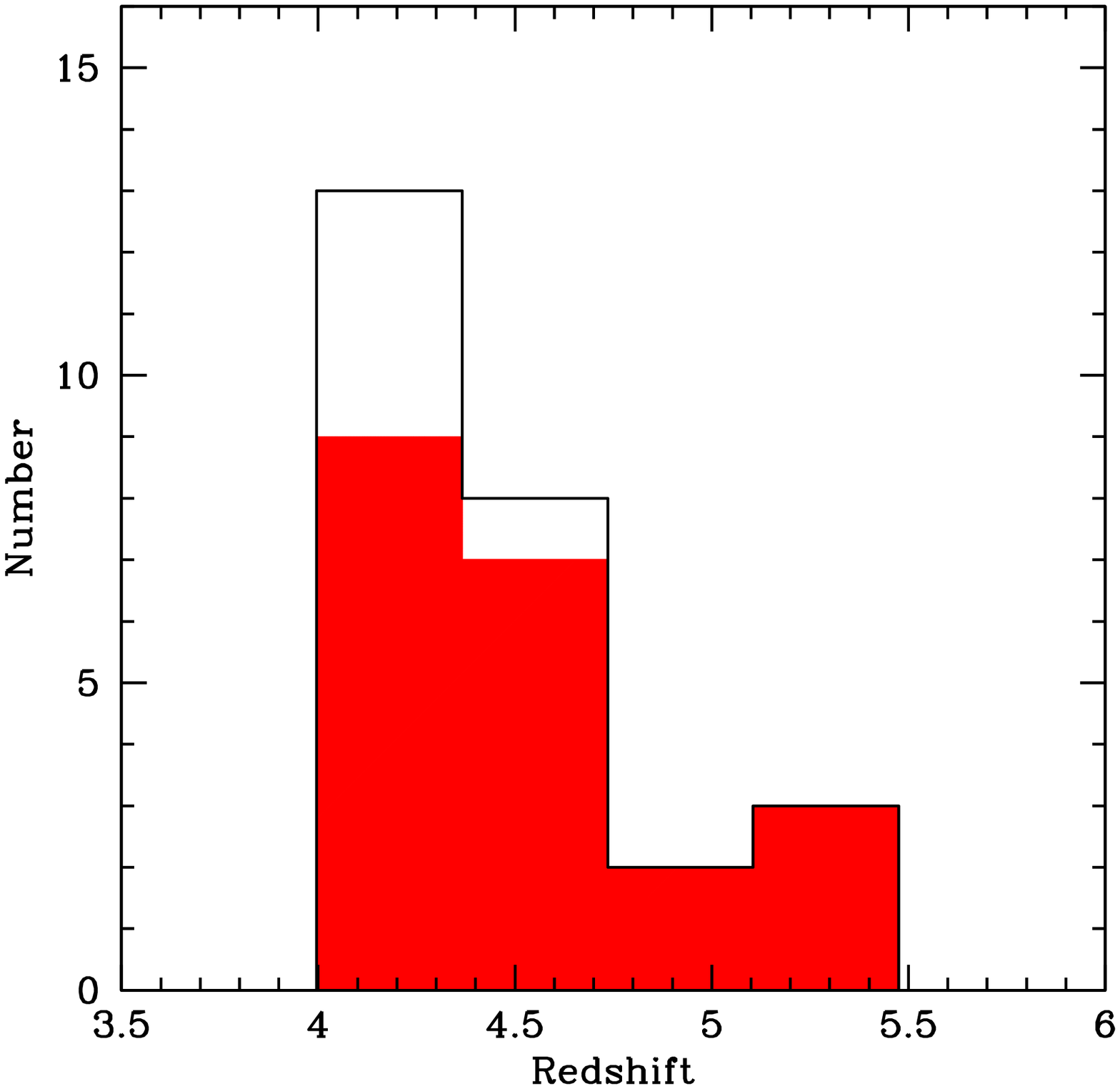}
   \caption{Redshift distribution of all the 26 z$>$4 QSO present in the CLASS survey. The shaded 
   area represents only those objects in the complete sample (see text for details)}
              \label{z_dist}
    \end{figure}
%----------------------------------------------------------------------------------------------

\section{Completeness of the sample}

As described in the previous section, the identification level of the sample
is very high (95\%). It is possible, however, that some high-z objects are missed
by our selection.  
To check the completeness of our sample we have searched through the
literature and in the SDSS spectroscopic database for high-z AGN among all the $\sim$11,000 CLASS sources, independently from 
their photometric colour, magnitude or sky position. We found 6 additional z$>$4 QSO 
(see Table~\ref{add}). 
Two of them were not selected because they have a low Galactic latitude ($|b^{II}|<$20 deg) while other
3 objects have a magnitude fainter than the adopted limit\footnote{One of the objects
  (GB6J160608+312504) presents also a large optical-to-radio offset
  (1.22$\arcsec$),
  significantly larger than typical radio (and optical) error uncertainty. In addition,
  an optical spectrum of this source is not available in the literature and in the
  CGRaBS catalogue (\citet{Healey2008}) the redshift is given as uncertain.
  We therefore consider this identification unreliable}. Only
one source (GB6J153533+025419) is not
recovered because it does not show a significant dropout ({\it g$-$r}=0.79 $\pm$0.07) and, therefore, it
can be considered as missed from our selection\footnote{In the SDSS this source has a higher 
dropout value ({\it g$-$r}=1.18) that would have allowed its inclusion in the sample. This suggests
that the low dropout value observed in PS1 can be related to some photometric problem in PanSTARRS}. 
This confirms that our selection method is highly complete for z$>$4 since it recovers nearly all (13/14) 
the already known high-z AGN in the sample. 
We note, instead, that for slightly lower values of redshift, the completeness level drops significantly: 
in Fig.~\ref{complet_log} we show the fraction of high-z AGN (with mag$<$21) from the literature or from SDSS DR12
that are present in CLASS (and in the sky area considered for the complete sample) and that are correctly recovered by our selection.
While most (95\%) of the z$>$4 objects are recovered, for redshifts below 4 this fraction decreases
significantly, being around 25\% for  3.6$<$z$<$4 and below 5\% for 3$<$z$<$3.6. 

We define as
{\it complete sample}, the one defined at high Galactic latitude ($|b^{II}|>$20 deg) and at a magnitude limit of 21 
(in {\it r}, {\it i} or {\it z} filter depending on the redshift of the source, as described in Section~2). This sample is composed by the 20 z$\geq$4 of Table~\ref{dropout_sample} plus
the additional object (GB6J153533+025419) present in Table~\ref{add}.  

\subsection{Impact of variability on the sample completeness}
Blazars are variable sources. Variability can potentially affect the completeness of the CLASS sample which was built using non-simultaneous radio data. Variability can in fact make an intrinsically flat spectrum variable source to appear as a  non-flat spectrum object depending on when fluxes have been measured in relation to the activity phase.  

 In order to estimate the importance of variability at radio frequencies we have first
compared the flux densities at 1.4~GHz computed in the NVSS and in the 
Faint Images of the Radio Sky at Twenty-cm (FIRST, \citealt{Becker1995}) surveys that have been
carried out at different epochs. Since all
CLASS objects at z$>$4 are compact at both FIRST and NVSS resolution, any observed difference in the flux densities
can be attributed to source variability and not to the different instrumental resolutions. 
From this comparison we have estimated a rms flux variability of 
$\sim$14\%.
We then used numerical simulations to evaluate the impact
of this variabilty on the final completeness of the sample. 
Assuming a flat distribution of 
radio spectral indices between -0.5 and +0.5  we found that about 6\% of intrinsically flat-spectrum sources could have been missed due to variability, i.e. $\sim$1-2 objects.
Using a gaussian distribution of the spectral indices centered at $\alpha$=0 would produce an even smaller percentage of missing sources due to variability.

An independent way to estimate this effect is to use the
flux densities available at other frequencies for the CLASS sources to derive independent estimates of the (non-simultaneous) spectral indices and to count the fraction of sources with $\alpha>$0.5.
Using the spectral indices between 150~MHz and 1.4~GHz,
described in Section~5.1, we have 1 object 
with $\alpha>$0.5 (See Fig.~\ref{class_alpha}). Using, instead, the indices between
1.4~GHz and 8.4~GHz, we have 2 sources (among the 
confirmed blazars, see discussion in the next section) with $\alpha>$0.5.
This is fully consistent with the results of our simulations.
Even if small, we will consider the impact of this potential incompleteness on the derivation of the space densities that will be presented in Section~6.  

Variability could, in principle, have an impact also on the optical selection of the
high-z candidates since different levels of non-thermal radiation from the beamed
jet can modify the optical colours of the source. However, unlike featureless blazars, FSRQ
typically have most of the UV/optical emission produced by the accretion disk/broad line
region while the emission from the relativistic jet is relatively less important at these
wavelengths. This
seems to be particularly true for high-z blazars, as discussed, for instance by \citet{Ghisellini2010}.
Therefore, we expect that variability in the optical band has an even lower impact on the
completeness of the CLASS sample compared to that at radio wavelengths.

%--------------------------------------------------------------------------------------
  \begin{figure}
    \centering
    \includegraphics[width=6cm, angle=0]{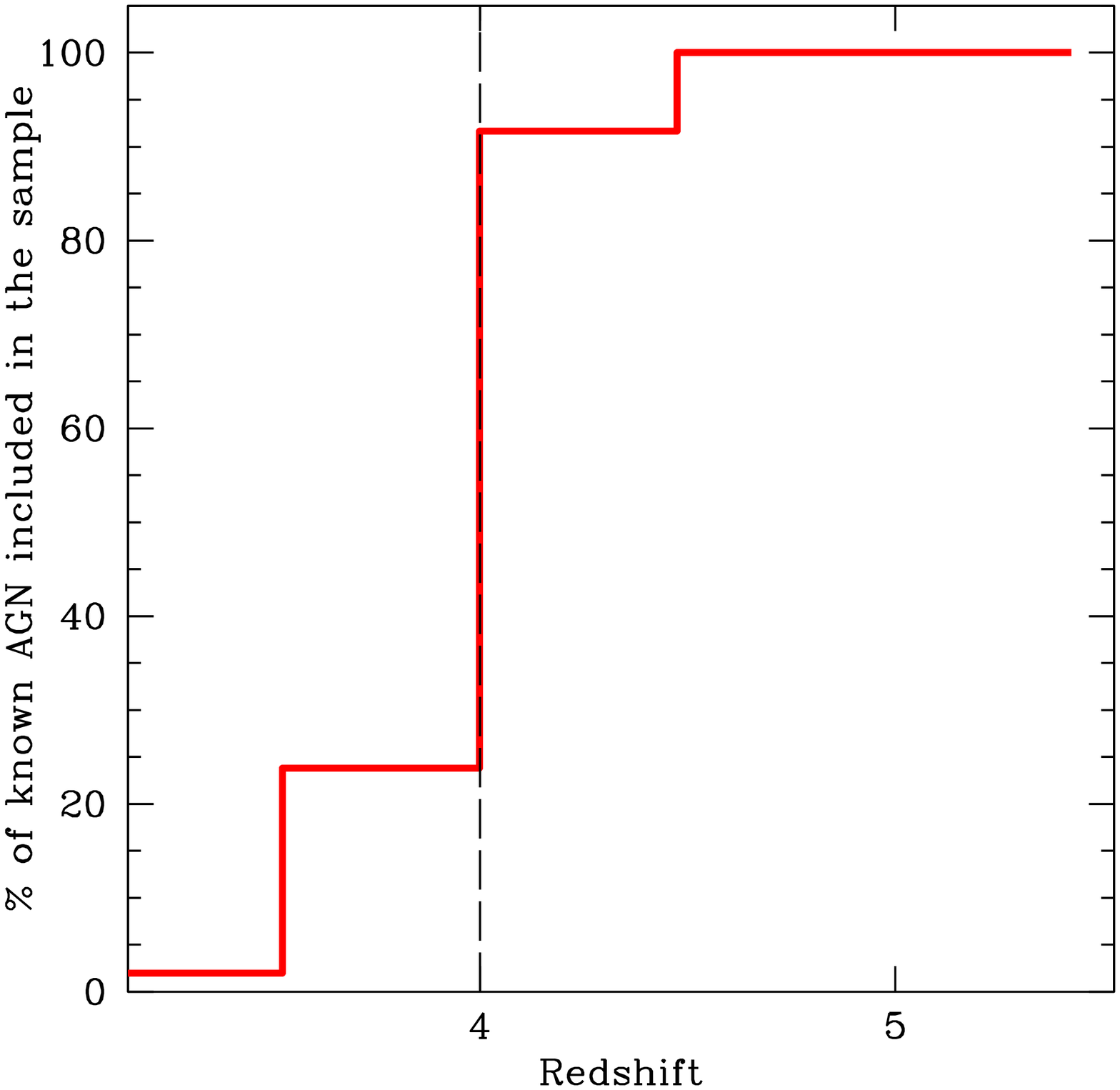}
   \caption{Fraction of know radio-loud AGN (from the literature or from the SDSS) that are recovered by our selection versus redshift. We are considering only the objects that fall in the area of sky that
has been used to define the CLASS sample.}
              \label{complet_log}
    \end{figure}
%-----------------------------------------------------------------------------------------------

\section{Radio spectral shapes: GPS vs flat-spectrum sources}
By definition, CLASS contains sources with a flat radio spectrum between 1.4 and 5 GHz (observed frame). 
For this reason, CLASS should preferentially select blazars although sources that do not have a genuinely flat 
spectrum across the entire radio band could be also included. 

The main contamination is that from AGN whose radio spectrum peaks at high-frequencies.
These objects, usually 
called Giga-Hertz Peaked Spectrum (GPS) sources (\citealt{ODea1998}), appear compact at arcsecond resolution 
and with an apparent flat radio spectrum simply because we are observing the self-absorbed part 
of the radio spectrum. These are intrinsically small radio-sources, possibly because of their young age. 
Interestingly, radio-loud AGN at high redshift
often show these characteristics (\citealt{Frey2003}; \citealt{Frey2005}; \citealt{Coppejans2016a}; 
\citealt{Coppejans2017}).
At m.a.s. scales the emission is usually resolved, often showing a
Compact-Symmetric Objects (CSO) morphology, i.e. a small scale version of a radio-galaxy
with lobes, hot-spots, jets and relatively weak cores.
This is an indication that the relativistic jet is mis-aligned
with respect to the observer. Examples are J2102+6015, at z=4.57
(\citealt{Frey2018}), J1606+3124, at z=4.56 (\citealt{Coppejans2017}) and  
J1427+3312, at z=6.12 (\citealt{Frey2008}, \citealt{Coppejans2016a}).
It is important to note, however, that
this is not always true: the source J0906+6930 at z=5.47 shows a clearly peaked radio spectrum (\citealt{Coppejans2017}) 
with a turnover frequency of 6.4~GHz corresponding to 41.4~GHz in the source rest-frame. At the same time,  
Very Long Baseline Interferometry (VLBI) data found evidence of Doppler boosting (\citealt{Zhang2017}, \citealt{Frey2018}), which was further supported by the study of radio variabilty (\citealt{Liodakis2018}) and of the SED (\citealt{An2018}). These results suggest that J0906+6930 is
a GPS source likely oriented towards the observer. Indeed, this is considered the highest redshift blazar discovered
so far (\citealt{Romani2004}). Therefore, the simple observation of a peaked radio spectrum does not firmly 
exclude the presence of beaming. 

In order to distinguish between truly flat spectrum sources from GPS it is important to 
extend the analysis of the radio spectrum at higher and lower
frequencies to detect possible hints of spectral curvature. 
If a source is a GPS we expect that the radio spectrum, sampled in a broad
range of frequencies, will not be well represented by a single power-law.
In this case we expect that the two-point spectral index computed at very high
frequencies would appear steeper than the one computed at lower frequencies (convex spectrum).

The analysis of the radio spectral shape of
CLASS high-z sources is presented in the following section.

%______________________________________________ Gamma_1 (lg rho, lg e)
   \begin{figure}
   \centering
    \includegraphics[width=8cm, angle=0]{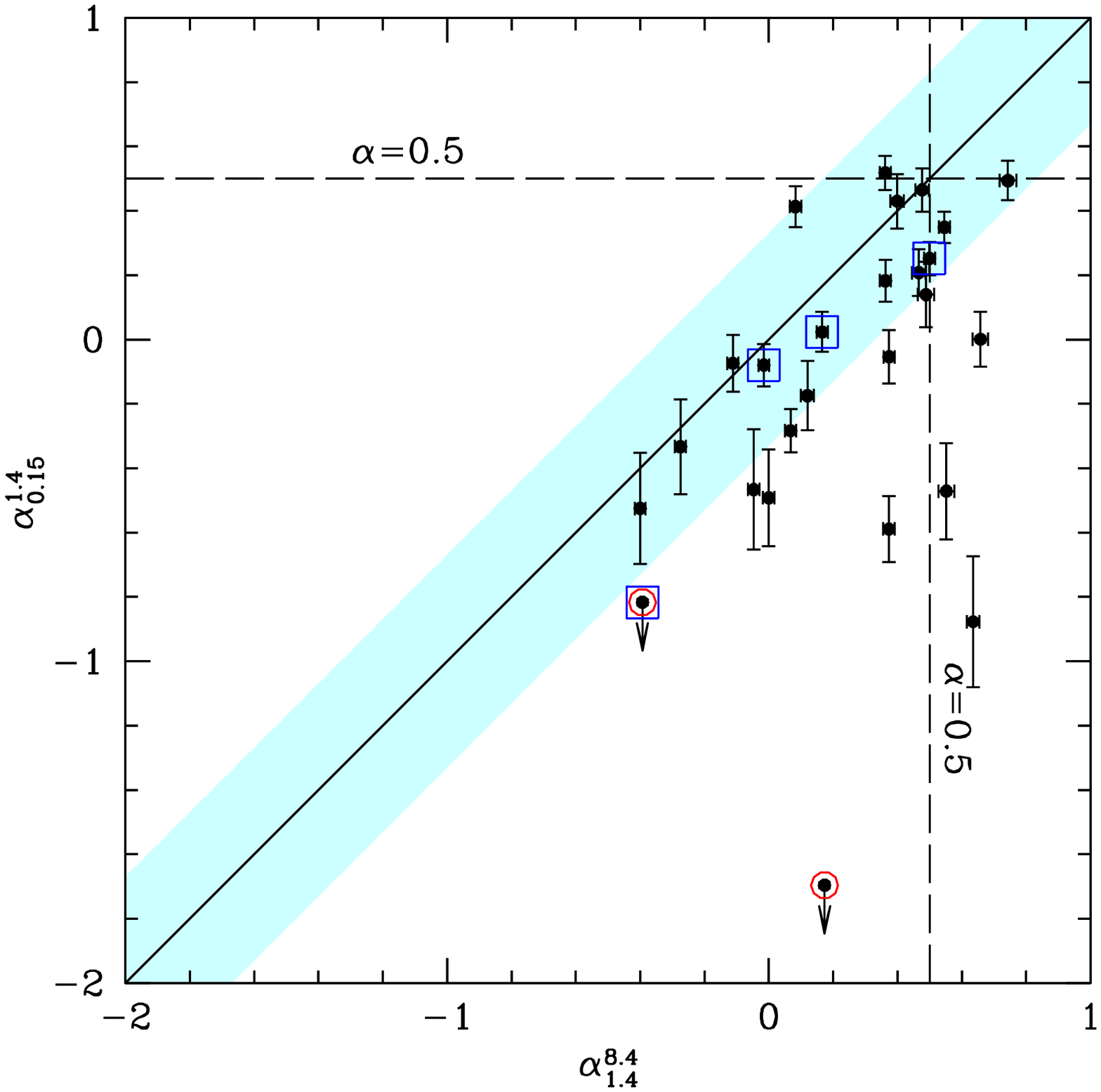}
   \caption{Radio spectral indices between 150~MHz and 1.4~GHz versus  
spectral indices between 1.4~GHz and 8.4~GHz of the AGN in CLASS with z$>$4
(from Table~\ref{dropout_sample} and Table~\ref{other}). 
Arrows indicate upper limits on $\alpha_{0.15}^{1.4}$  (i.e. sources not detected 
in the TGSS at 150~MHz). The boxes mark the objects
already confirmed as blazars in the literature while the red open circles mark
the sources (GB6J090631+693027, GB6J160608+312504) classified as peaked radio spectrum by 
\citet{Coppejans2017}.
The continuous line is the 1:1 relation while the light-blue shaded area indicates the expected spread
(90\% confidence level) due to variability (see text for details).  
}
              \label{class_alpha}
    \end{figure}
%-------------------------------------------------

\subsection{Low-frequency data from TGSS}
In order to cover the widest range of frequency, we have used the recently released TGSS survey 
(TIFR GMRT Sky Survey, \citealt{Intema2017}) which is 
carried out at 150~MHz with the Giant Metrewave Radio Telescope (GMRT), and the high frequency
data at 8.4~GHz, taken at VLA in A-configuration, 
that are available for the CLASS objects.
For the range of redshift considered here, the TGSS
survey covers the source rest-frame frequency of 0.85~GHz (at the average
redshift of the sample, z=4.5) while the 1.4~GHz and 8.4~GHz flux densities
correspond, respectively, to  8~GHz and 46~GHz.
Therefore, a spectral index computed 
by combining TGSS data with flux densities at 1.4~GHz (from NVSS) gives us indications of the spectral
slope between $\sim$0.85 and 8~GHz i.e. in a range similar to that considered 
using NVSS and GB6 data  for local sources.  The indices computed between 1.4~GHz and 8.4~GHz, instead, correspond to a high frequency (8-46~GHz) range.

The current data release of TGSS (ADR1) 
covers 90\% of the sky (36,900 deg$^{2}$ at declination greater than -53$\deg$)
at an rms of 5~mJy beam$^{-1}$ and with a resolution of 25\arcsec$\times$25\arcsec (in the northern hemisphere). 
Out of the 26 high-z sources in the CLASS complete sample, 17  are present in the TGSS catalogue.
For the remaining objects we have analysed the TGSS images and looked for a detection at lower significance.
We found a detection (at least 2$\sigma$) in 7 out of the 9 objects not included in the TGSS source catalogue.
For the two undetected objects we consider an upper limit of the 150~MHz flux density
of 15~mJy = (3$\times$rms). 

In Fig.~\ref{class_alpha} we show the spectral indices between 150~MHz and 1.4~GHz compared to the
indices between 1.4~GHz and 8.4~GHz.  
It is important to note that these spectral indices are computed using non-simultaneous
data. Variability can affect the spectral classification into blazars and GPS-like objects as shown, for instance, by \citet{Orienti2007a}.
On Fig.~\ref{class_alpha} we plot the expected spread (90\% confidence level) 
on the 1:1 relation assuming an rms variability of 14\%, estimated in the previous section, and assuming that 
variability is similar at all frequencies.
Sources that are distant from the shaded area have a spectrum that is likely not well represented by a single power-law.

From Fig.~\ref{class_alpha} we see that 18 objects 
 fall within the shaded area around the 1:1 relation (or they are consistent
with it considering the error bars) thus suggesting that their
spectrum is well represented by a flat power-law in a wide range of frequencies (between 150~MHz up to 8.4~GHz, corresponding to $\sim$1-50~GHz rest-frame),
without significant curvatures. 
 
In 8 cases, instead, the  
$\alpha_{0.15}^{1.4}$ is significantly flatter than
$\alpha_{1.4}^{8.4}$, typical of GPS-like sources (convex spectrum). 
The most striking one is GB6J160608+312504 whose limit on $\alpha_{0.15}^{1.4}$
is significantly different from the slope measured
between 1.4~GHz and 8.4~GHz. 
This object is known to have a spectrum peaking at 2.6~GHz (\citealt{Coppejans2017}),
corresponding to 14.4~GHz in the source rest-frame. VLBI observations suggest a 
Compact-Symmetric Objects (CSO) morphology (\citealt{Coppejans2017}). 
Another object that lies below the 1:1 relation in Fig.~\ref{class_alpha} is GB6J090631+693027, 
which, again, has been classified by \citet{Coppejans2017} as GPS (see discussion above).

In Tab.~\ref{table_radio_spectra} we report the radio spectral type (``Flat''/''Peaked'') as derived from the analysis of Fig.~\ref{class_alpha}.  

In summary, 18 out of 26 high-z sources in CLASS (14 in the complete sample) have a spectral index between 0.15 and 1.4 GHz
consistent with the index between 1.4 and 8.4~GHz (considering the possible variability) i.e. they 
have a genuine flat radio spectrum  on a wide range of frequencies, between 150~MHz up to 8.4~GHz, 
corresponding to 0.83-46~GHz in the sources' rest-frame (at the
average redshift of the sample). This strongly supports the idea that these are
beaming dominated objects and, therefore, likely observed at relatively small angles. 

For the remaining objects the situation is less clear since
we have indications for a significant change of shape
which may suggest a GPS nature. These sources could be mis-aligned AGN 
although we cannot completely exclude their blazar nature, as in the case
of J0906+6930 (GB6J090631+693027).

%%-------------------------------------------------------
   \begin{figure*}
   \centering
\includegraphics[width=15cm, angle=0]{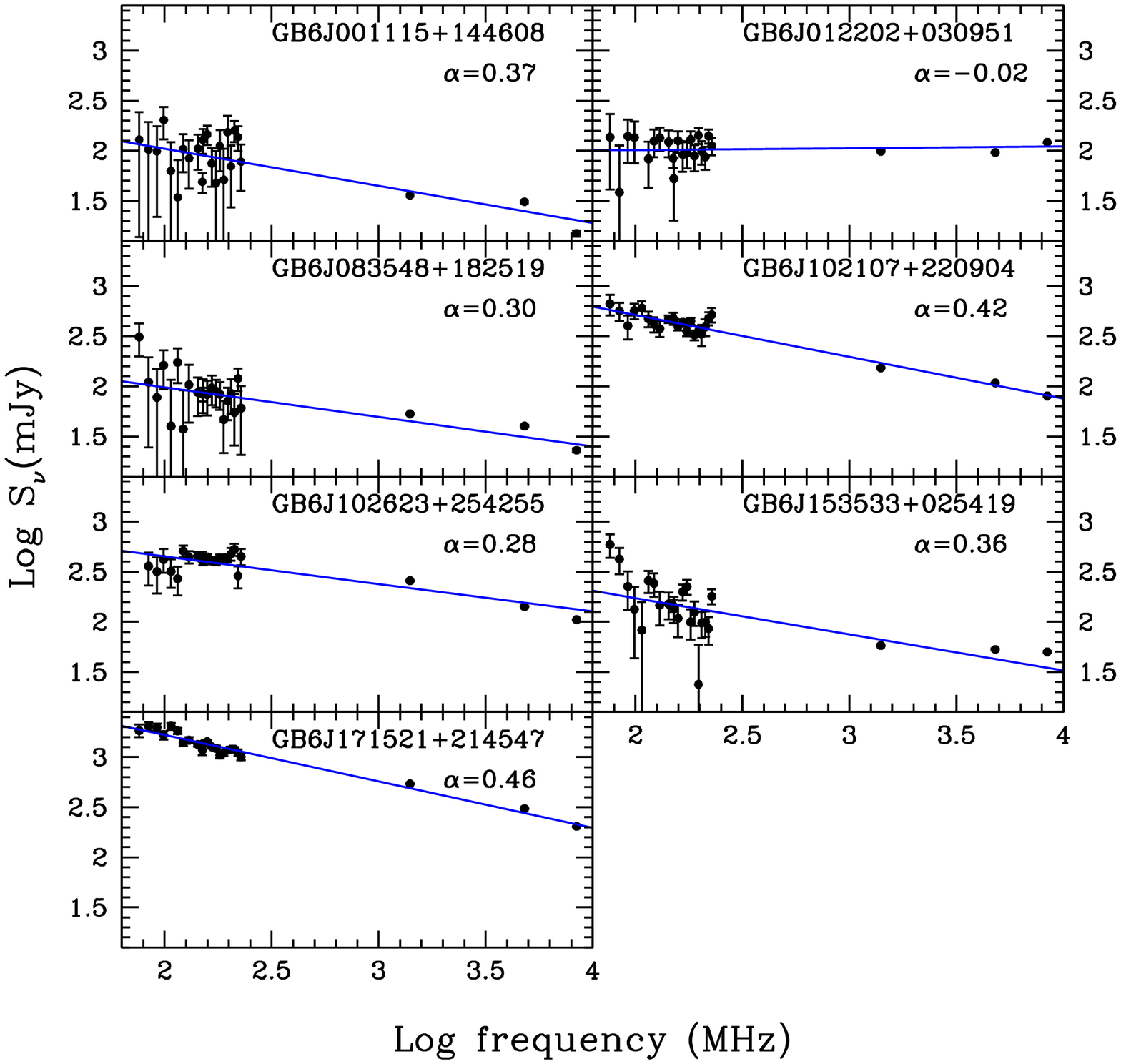}

   \caption{Radio spectra (observed frame) of the 7 high-z from CLASS detected in the
GLEAM survey}
              \label{all_radio_spectra}
    \end{figure*}
%-------------------------------------------------

\begin{table*}
\caption{Spectral radio properties}
\begin{tabular}{l r r r r r l}
\hline\hline
 name & S$_{0.15GHz}$ & S$_{1.4GHz}$ & S$_{8.4GHz}$ & $\alpha_{0.15}^{1.4}$ & $\alpha_{1.4}^{8.4}$ & Spectral type\\ 
      & (mJy)        & (mJy)       & (mJy)      &                          &                          &              \\
 \hline
{\bf  GB6J001115+144608 } &    49 &     36 &     15 &  0.14 &          0.49 & Flat:      \\
{\bf  GB6J003126+150729 } &    13 &     42 &     86 &  -0.53 &         -0.40 & Flat       \\
{\bf  GB6J012126+034646 } &    11 &     78 &     25 &  -0.88 &          0.64 & Peaked     \\
{\bf  GB6J012202+030951 } &    84 &     99 &    121 &  -0.07 &         -0.11 & Flat       \\
  GB6J025758+433837   &    79 &    149 &    132 &  -0.28 &          0.07 & Flat:      \\
{\bf  GB6J083548+182519 } &    84 &     53 &     23 &  0.21 &          0.47 & Flat       \\
{\bf  GB6J083945+511206 } &    15 &     43 &     16 &  -0.47 &          0.55 & Peaked     \\
{\bf  GB6J090631+693027 } & - &     93 &    188 & $<$-0.82 &         -0.39 & Peaked     \\
{\bf  GB6J091825+063722 } &    21 &     31 &     25 &  -0.17 &          0.12 & Flat       \\
  GB6J102107+220904   &   486 &    153 &     80 &  0.52 &          0.36 & Flat       \\
{\bf  GB6J102623+254255 } &   451 &    257 &    105 &  0.25 &          0.50 & Flat       \\
{\bf  GB6J132512+112338 } &    22 &     82 &     42 &  -0.59 &          0.37 & Peaked     \\
{\bf  GB6J134811+193520 } &    52 &     52 &     16 &  0.00 &          0.66 & Peaked     \\
{\bf  GB6J141212+062408 } &   123 &     47 &     23 &  0.43 &          0.40 & Flat       \\
{\bf  GB6J143023+420450 } &   177 &    211 &    217 &  -0.08 &         -0.02 & Flat       \\
{\bf  GB6J151002+570256 } &   213 &    202 &    150 &  0.02 &          0.17 & Flat       \\
{\bf  GB6J153533+025419 } &   146 &     58 &     50 &  0.41 &          0.08 & Flat       \\
  GB6J160608+312504   & - &    663 &    486 & $<$-1.70 &          0.17 & Peaked     \\
{\bf  GB6J161216+470311 } &   153 &     54 &     23 &  0.46 &          0.48 & Flat       \\
{\bf  GB6J162956+095959 } &   160 &     53 &     14 &  0.49 &          0.74 & Flat       \\
{\bf  GB6J164856+460341 } &    12 &     34 &     37 &  -0.47 &         -0.05 & Flat:      \\
{\bf  GB6J171103+383016 } &    15 &     45 &     45 &  -0.49 &          0.00 & Peaked:    \\
  GB6J171521+214547   &  1177 &    540 &    203 &  0.35 &          0.55 & Flat       \\
  GB6J195135+013442   &    47 &     99 &    162 &  -0.33 &         -0.27 & Flat       \\
{\bf  GB6J231449+020146 } &   111 &    125 &     64 &  -0.05 &          0.37 & Peaked:    \\
{\bf  GB6J235758+140205 } &   167 &    111 &     58 &  0.18 &          0.36 & Flat       \\
\hline
\end{tabular}

{\bf column 1}: name (in bold-face the sources belonging to the complete sample); {\bf column 2}: flux density at 150~MHz from TGSS; {\bf column 3}: flux density at 1.4~GHz from NVSS; {\bf column 4}: flux density at 8.4~GHz from CLASS;
{\bf column 5}: spectral index between 150~MHz and 1.4~GHz; {\bf column 6}: spectral index between 1.4~GHz and 8.4~GHz; {\bf column 7}: spectral type. 
The symbol ``:'' indicates that radio spectral classification is uncertain.
\label{table_radio_spectra}
\end{table*}

It is interesting to compare these numbers with the ones discussed in \citet{Coppejans2017} who 
studied the radio spectral shapes of a sample of 30 high-z (z$>$4.5) radio-loud AGN selected at 1.4~GHz. 
They grouped the sources into 3 classes: flat-spectrum (blazar) objects, steep-spectrum objects
and sources with a peaked radio spectrum. They found that the three classes are almost evenly populated.
In CLASS, 
where the selection  is made at
higher frequencies (5~GHz) and where steep spectrum objects have been 
excluded by definition, we found that the majority (at least 69\%) of the objects are truly flat-spectrum object 
over a wide range of frequencies.
This result supports the strategy of using a 5~GHz selection, coupled to the constraint on the spectral slope
between 1.4 and 5~GHz, to efficiently select a sample of high-z blazars. In contrast, in a radio selection 
carried out at lower frequencies (e.g. 1.4~GHz) and without any constraint on
the radio slope, 
the blazar selection efficiency falls down to $\sim$30\%. 

\subsection{Low-frequency data from GLEAM}
In the analysis presented in the previous section we have 
inferred the radio spectral shapes on the basis of 
the comparison between the slopes computed between two 
frequencies. 
This approach was imposed by the limited amount of data points available for each object.  
In some cases, however, we can 
carry out a proper spectral analysis thanks to the
availability of several (simultaneous) data points taken from 
the GaLactic and Extragalactic All-sky MWA (GLEAM) survey (\citealt{Hurley-Walker2016}). GLEAM is a low-frequency
radio survey carried out with Murchison Widefield Array (MWA), 
the low-frequency Square Kilometre Array (SKA1 LOW) precursor located in Western Australia. 
The first release of the extragalactic catalogue covers
the entire high Galactic latitude ($>$30$^\circ$) sky south of $+30^\circ$ of declination (24,831 square degrees),
excluding some areas such as the Magellanic Clouds. The catalogue contains flux density measurements 
in 20 separate intervals between 72 and 231~MHz with a flux density limit (90\% completeness) of 170~mJy. It should be noted, however, that the limited resolution ($\sim$ 100$\arcsec$) of the telescope may affect the
spectral analysis if strong, nearby sources are present.

In total, 7 CLASS high-z sources are currently detected in GLEAM, out of the 16 objects falling in the sky 
area covered by the survey. In Fig.~\ref{all_radio_spectra} we show
the wide band radio spectra, from 72~MHz up to 8.4~GHz, of these seven sources. In particular, we report
the integrated flux densities from the GLEAM catalogue. We fitted all the photometric points with a simple
power-law to check the actual radio spectral slopes.  Overall, the radio spectra computed on a wide range 
of frequencies are flat ($\alpha<$0.5), confirming the classification
previously derived using TGSS and high-frequency data. Only in one case (GB6J153533+025419)
there are hints for a steepening at low frequencies (below 80-100~MHz, corresponding to 400-550~MHz in the
sources'rest frame) probably indicating the presence of extended, non-beamed radio emission. 

Overall, the spectral analysis of the few CLASS objects with data
from GLEAM supports the results obtained in the previous section, based on few, non simultaneous, data points giving us confidence
on the reliability of our radio spectral classification.

\section{Space density of high redshift blazars}
Thanks to its completeness the CLASS sample can be used to 
derive the space density of blazars above z=4. In particular,
we can compute the space density of blazars with a radio
power at 5~GHz between 10$^{27}$-1.3$\times$10$^{28}$ W Hz$^{-1}$, 
which is the observed range of luminosities. 
The complete sample contains 21 objects with redshift between 4 and 5.4. Fourteen have been classified as blazars, as discussed  in the previous sections\footnote{For consistency, we are not considering GB6J090631+693027 as blazar, given its peaked radio spectrum}. Since 
we cannot exclude that some of the remaining 7 objects are blazars, we 
consider the total sample as an upper limit on the number of blazars.
The CLASS survey covers 16 300 deg$^2$ of sky but we are now considering 
the complete sample that is defined for 
Galactic latitudes above 20 deg (see Section~2) corresponding to a sky area  of 
13 120 deg$^2$ (= 4 steradians) i.e. 1/3 of the total sky area. 

In order to compute the space density of high-z blazars we use the method described,
for instance, by \citet{Avni1980}: for each object with redshift between z$_1$ and
z$_2$ we compute the
volume of the Universe within which it could have been discovered as: 

\begin{center}

$V_{obs} = \frac{A}{4\pi} V_{max}$
  
\end{center}

where A is the sky area (in steradians) covered by the survey and $V_{max}$ is the co-moving
volume of Universe between $z_1$ and min($z_2$, $z_{max}^{R}$, $z_{max}^O$). The quantities 
$z_{max}^R$ and $z_{max}^O$ are, respectively,  
the redshift at which the observed flux from the source would be equal to the radio/optical flux limit of the survey.
The reason for taking the minimum of the 3 values of redshift is that the $V_{max}$ must be computed on the most
stringent among the radio/optical constraints. If both values are larger than z$_2$ then the V$_{max}$ is simply
the co-moving volume of Universe between z$_1$ and z$_2$.

The space density then is:

\begin{center}

 $ \rho = \sum \frac{1}{V_{obs}}$

\end{center}

with an error that is given by:

\begin{center}

$ err_{\rho} = \frac{err}{N} \rho$

\end{center}

where N is the observed number of blazars in the given redshift bin and $err$ is
the Poissonian uncertainty on N.

Using this method\footnote{for the k-corrections we have assumed an optical spectral index of 0.44 (\citealt{VandenBerk2001}) and we used the radio spectral index computed between 150~MHz and 1.4GHz. We note that the assumption on the optical spectral index has a very little impact on the final results since we are considering a very limited rest-frame wavelength range. A variation of the optical spectral index of $\pm$0.5 only produces a variation smaller than $\pm$0.04 dex on the computed space densities}, we compute a space density of blazars at redshift between 4 and 5.5 and with radio powers at 5~GHz between  10$^{27}$ and 1.3$\times$10$^{28}$ W Hz$^{-1}$ of $\rho_{blaz}$=0.13$^{+0.05}_{-0.03}$ Gpc$^{-3}$. 
Considering all the objects of the sample the density is $\rho_{all}$=0.17$^{+0.05}_{-0.04}$ Gpc$^{-3}$. 
This is the first actual estimate of the blazar space density in this 
range of redshift.

As explained in the previous sections, there are two potential sources of incompleteness, one related to the lack of spectroscopic confirmation of two high-z candidates (see Section~3) and one due to variability (see Section~4.1). 
In particular, one of the two candidates still unclassified (GB6J064057+671228) is potentially interesting since it has a flat radio
spectrum (including the flux density at 150~MHz) and it is an {\it r}-dropout source i.e. likely at z$>$4.5.
The other one (GB6J061110+721814), instead, is a {\it g}-dropout (i.e. with redshift between 3.5 and 4.5) and
with a possibly peaked radio spectrum. If we take these objects as true high-z sources, 
and if we consider also the possible fraction of missing objects due to variability, we
obtain marginally higher space densities ($\rho_{blaz}$=0.15 Gpc$^{-3}$ 
and $\rho_{all}$=0.19 Gpc$^{-3}$).

\section{Discussion}
It is now interesting to compare the space densities computed in the previous 
section with the predictions based on the 
radio selected sample of FSRQ recently published by \citet{Mao2017} (hereafter M17).  
The sample is based on the combination of FIRST and GB6 surveys and it has
a relatively high radio flux limit (220 mJy) and covers 9500 deg$^{2}$ of
sky. Given the large flux limit, this sample mostly contains objects with
redshift below 4, with only one source at redshift above 4 (GB6J102623+254255, 
z=5.28). Based on these data, M17 have found that the best
parametrization for the cosmological evolution of FSRQ is a luminosity-dependent
density evolution (LDDE) where the space density of FSRQ peaks at a redshift
that depends on the
luminosity of the object, being higher for the objects with higher
radio luminosities. The radio luminosity function, instead, is well
represented by a double power law function. We used the analytical
form presented in M17 to derive the expected space density as
a function of redshift of the FSRQ in the same luminosity range of the
high-z objects in CLASS. Since M17 defined their sample at 1.4~GHz
we compute the range of radio luminosities of CLASS high-z objects at this
frequency (Log L$_{1.4~GHz}$=P$_{1.4 GHz}\times\nu_{1.4 GHz}$=43.04-44.43 erg s$^{-1}$). 
We then integrated the radio
luminosity function of M17 in this range of luminosities
using the best-fit LDDE evolution.
The result is reported in Fig.~\ref{figure_dens} together with 
the densities computed using the CLASS complete sample in two bins of 
redshifts. We also added the upper limit between z=5.5 and z=6 
based on the fact that the none of the candidates in this range of redshift (i.e.
the objects with a dropout in the {\it i}-band)
turned out to be true high-z objects.

The CLASS data points 
follow the predictions of M17 that are based on sources with
redshift much lower than the ones of the CLASS objects. 
Even considering the missing identification in CLASS (GB6J064057+671228),
which could be a z$>$4.5 blazar, the densities computed with the
CLASS sample are still consistent with the 
predictions of M17.

According to this modelization, the peak of the
space density, in this luminosity bin, is at z$\sim$2, a value similar
to that found for the radio-quiet QSO (e.g. \citealt{Hopkins2006}).
This is significantly different from the evolution estimated by 
\citet{Ajello2009} using an
X-ray selected sample of blazars derived from the Swift-BAT survey that shows a peak
at much higher redshifts z$\sim$4 (\citealt{Ajello2009}). 
The difference cannot be ascribed to a different redshift sensitivity of the two samples since 
the redshift interval of the objects used by \citet{Ajello2009} is similar to that
of the FSRQ considered by M17.

To further evaluate the position of the peak in the space density distribution in this range
of radio luminosity, we have considered the QSO in CLASS with a redshift of $\sim$3. At this
redshift all the QSO with a radio luminosity in the considered range, have a flux density
above the limit of 30 mJy and, therefore, they are expected to be included in the sample. 
We consider the sky area covered by the SDSS to maximize the number of identifications.
In this area ($\sim$28\% of the total sky) the number of QSO in CLASS with z between 3 and 3.2 and
a (log) radio luminosity at 1.4~GHz between 43.04 and 44.43 erg s$^{-1}$ is 38 corresponding to 
a space density of 1.4 Gpc$^{-3}$. This is a lower limit since the identification level of
CLASS, even considering the area covered by SDSS, is relatively low ($\sim$15\%). 
However, we expect that most of the QSO in this redshift range have been already identified, at 
least down to the magnitude limit of SDSS, thanks to the dedicated searches carried out
in the past using the dropout method (in this case, the dropout falls in the u-band, \citealt{Fan1999}).
In any case, the lower limit (see Fig.~\ref{figure_dens}) is clearly much higher than the 2 points at z$>$4 thus
excluding the hypothesis of a peak at z$\sim$4. 
 
The observed discrepancy between the results based on radio selected samples and those from
the BAT survey  needs to be investigated. In the context of a luminosity-dependent density
evolution, the observed differences could be explained if the blazars selected by BAT
were systematically more luminous (in the radio band) than those selected by M17 or in the
CLASS survey. Indeed, about 80\% of the blazars in the BAT sample with z$>$2 (i.e. those that
allow to constrain the peak of the evolution) have a radio luminosity above  10$^{44}$ erg s$^{-1}$
while only 14\% of the CLASS high-z objects in the complete sample have a radio luminosity in this range.
It is thus possible that the space density of blazars with the highest radio luminosities,
above the range observed in CLASS, peaks at larger redshift than observed at lower luminosities. 
If confirmed, this would imply that only the most luminous radio-loud QSO have a cosmological 
evolution that peaks at very high redshifts ($\sim$4) the majority having an evolutionary behaviour very similar to that observed in radio-quiet QSO, peaking at z$\sim$2. 

Another possibility is that radio and X-ray selections 
are sampling different classes of objects having different cosmological behaviours. 
This is not unusual within the blazar family: BL Lac objects are known
to show significantly different cosmological evolution depending on the selection band, being
weakly positive in radio selections (\citealt{Stickel1991}; \citealt{Marcha2013}), negative 
in X-ray selections (e.g. \citealt{Morris1991a}; \citealt{Rector2000}; \citealt{Beckmann2003})
and showing little or no evolution when mixed (radio plus X-rays) selections
are used (e.g. \citealt{Caccianiga2002b}). One of the discussed possibilities is that the two 
selections are sampling different 
sub-populations of BL Lacs (the so-called High-Energy peaked BL Lac and Low-Energy peaked
BL Lac) that evolve in a different way with cosmic time. 
A similar difference could be present also among the class of FSRQ:  
in this case, X-rays (BAT) and radio surveys may be sampling
different sub-populations of FSRQ having different X-ray-to-radio luminosity ratios and
different cosmological evolutions.   
A systematic study of the X-ray properties of all the high-z
CLASS objects is in progress in order to establish the difference/similarities 
between this
sample and the blazars selected in the X-rays at lower redshifts (see. \citealt{Ighina2018} for first results of this analysis).

%______________________________________________ Gamma_1 (lg rho, lg e)
   \begin{figure}
   \centering
     \includegraphics[width=8cm, angle=0]{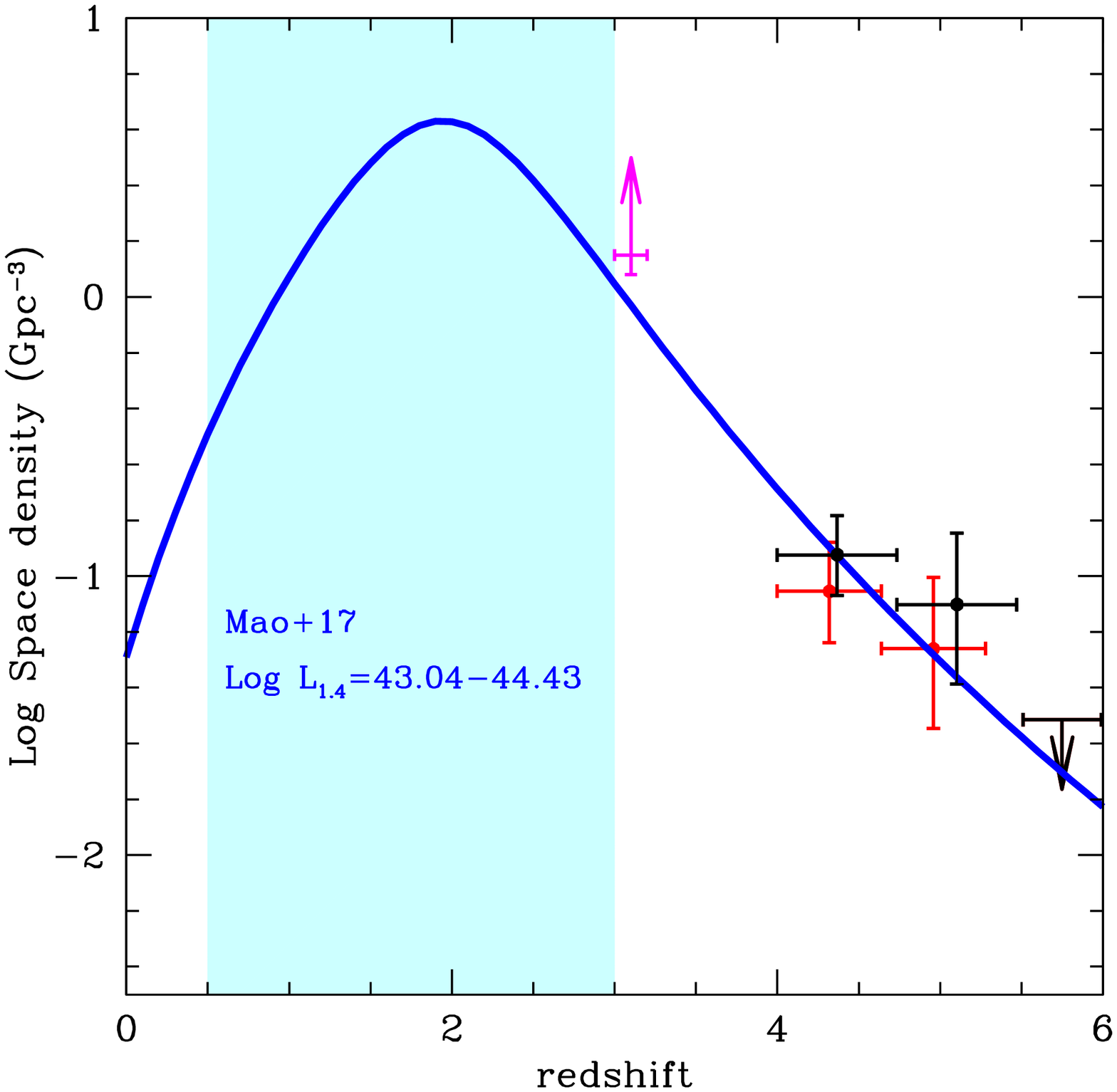}
   \caption{Space density versus redshift of the blazar candidates in CLASS considering 
all the 21 objects in the complete sample (black point) or just those with a 
flat radio spectrum in a wide range of frequencies (14 objects, red points). The blue solid line 
represents the  expected densities computed by integrating the best fit radio luminosity function 
of M17 in the same range of radio luminosities (at 1.4~GHz) observed in
the CLASS high-z sample and using the LDDE cosmological evolution found by M17
(clean sample). The shaded area indicates the range of redshift actually
sampled by the FSRQ studied by M17 in the indicated range of
radio luminosity. The magenta arrow represents the lower limit on the space density
derived from CLASS at z$\sim$3 (see text for details). 
}
              \label{figure_dens}
    \end{figure}
%-------------------------------------------------

\section{Summary and conclusions}
We have presented a well-defined, radio flux limited sample of 26 high-z ($>$4) blazar candidates selected from CLASS, which is a survey of flat-spectrum radio sources (between 1.4 and 5 GHz). 
In particular, twenty-one objects constitute a statistically complete sample
down to a magnitude of 21, thanks to dedicated spectroscopic follow-ups at LBT
and TNG. Notably, these observations have provided 2 new z$\geq$5 objects 
that represent a significant increase of the number of blazars
above z=5 currently known (4 objects found to date).

By studying the radio spectrum over a wide range of frequencies (from 150~MHz to 8.4~GHz,
corresponding to $\sim$0.85-46~GHz in the rest-frame of the sources) we found that more than half 
($\sim$18 sources) of the 
CLASS objects are genuine flat-spectrum sources, the remaining fraction being 
likely represented by GPS-like objects. 
Considering the complete sample, the number of truly flat spectrum sources is
14. We cannot exclude, however, that some of the objects with a GPS-like
radio spectrum could be blazars. High resolution radio data (VLBI) and/or variability studies are needed
in order to establish their blazar/non-blazar nature on a firm basis. 
We have an approved program at the European VLBI Network (EVN) to
carry out a complete follow up at 5~GHz all
the high-z sources in CLASS. 

We then used the complete sample to compute the space density of blazars at redshift 
between 4 and 5.5.  We found a value of  0.13$^{+0.05}_{-0.03}$ Gpc$^{-3}$, 
considering only the confirmed blazars, and 0.17$^{+0.05}_{-0.04}$ Gpc$^{-3}$, using the entire sample. Accounting for the possible incompleteness, the values of densities become 
$\rho_{blaz}$=0.15 Gpc$^{-3}$ 
and $\rho_{all}$=0.19 Gpc$^{-3}$, respectively.
This is the first estimate (not a lower limit or an extrapolation) 
of the space density
of blazars in this redshift interval.

The resulting space densities are in good agreement with the predictions recently presented by M17
using a sample of FSRQ with z$<$4. According to these authors, the evolution of FSRQ
of these luminosities should peak at a redshift $\sim$2, which is similar to what is observed
in radio-quiet QSO but significantly different from what has been found by \citet{Ajello2009} 
using an X-ray selected sample of blazars. The origin of this difference is not clear and it may
suggest that only the most luminous radio-loud QSO present a 
peak of the space density at very high redshift, while the bulk of the population has a cosmological evolution more similar to radio-quiet QSO.

\section*{Acknowledgments}
We thank I.W. Browne for useful discussions and the referee for his/her comments that improved the quality of the paper.
This work is based
on observations made with the Large Binocular Telescope (LBT,
program IT-2017B-006)
and the Telescopio Nazionale Galileo (TNG, programs A36TAC25 and A37TAC5).
In an early stage of the project, this work has also used data from 
the William Herschel Telescope (WHT, program SW2015a48) and
the Apache Point Observatory 3.5-meter Telescope.
LBT is an international collaboration
among institutions in the United States of America, Italy and
Germany. The LBT Corporation partners are the University of Arizona
on behalf of the Arizona university system and the Istituto
Nazionale di Astrofisica.
TNG is operated on the island of La Palma by the Fundaci\'on Galileo Galilei of the INAF 
(Istituto Nazionale di Astrofisica) at the Spanish Observatorio del Roque de los Muchachos of the 
Instituto de Astrofisica de Canarias
WHT is operated on the island of La Palma by the Isaac Newton Group (ING).
APO 3.5-meter telescope is owned and operated by the Astrophysical Research
Consortium.
This work is based on SDSS data. 
Funding for the Sloan Digital Sky Survey IV has 
been provided by the Alfred P. Sloan Foundation, the U.S. Department of Energy 
Office of Science, and the Participating Institutions. SDSS-IV acknowledges
support and resources from the Center for High-Performance Computing at
the University of Utah. The SDSS web site is www.sdss.org.
This publication makes use of data products from the Wide-field Infrared Survey Explorer, which is a joint project of the University of California, Los Angeles, and the Jet Propulsion Laboratory/California Institute of Technology, funded by the National Aeronautics and Space Administration. 
We thank the staff of the GMRT that made these observations possible. GMRT is run by the National Centre for Radio Astrophysics of the Tata Institute of Fundamental Research.
AC, AM, SB, RDC, PS acknowledge support from ASI under contracts ASI-INAF n. I/037/12/0 and n.2017-14-H.0  and from INAF under PRIN SKA/CTA ‘FORECaST’.
CC acknowledges funding from the European Union's Horizon 2020 research and 
innovation programme under the Marie Sklodowska-Curie grant agreement No 664931
SA acknowledges support from CIDMA strategic project (UID/MAT/04106/2013), 
ENGAGE SKA (POCI-01-0145-FEDER-022217), funded by COMPETE 2020
and FCT, Portugal.

\bibliographystyle{mn2e}
\bibliography{blazar_high_z}

\end{document}